%% file: ws-ijseke.tex
\documentclass{ws-ijseke}
\usepackage[numbers]{natbib}
\usepackage{rotating}
\usepackage{multirow}
\usepackage{booktabs}
\usepackage{indentfirst}          
\usepackage{float}

\usepackage[most]{tcolorbox} 
\tcbset{
  enhanced,
  boxrule=0.4pt,          
  colframe=black!40,      
  arc=1mm,                
  left=6pt,right=6pt,top=6pt,bottom=6pt,
  boxsep=4pt,
  before skip=3pt,
  after skip=3pt
}

\newtcolorbox{topbox}{
  colback=green!10,
  overlay={
    \node[anchor=south east,
          inner sep=1pt,
          font=\footnotesize\bfseries]
      at ([xshift=-1pt,yshift=1pt]frame.south east) {(a)};
  }
}

\newtcolorbox{bottombox}{
  colback=orange!10,
  overlay={
    \node[anchor=south east,
          inner sep=1pt,
          font=\footnotesize\bfseries]
      at ([xshift=-1pt,yshift=1pt]frame.south east) {(b)};
  }
}

\newtcolorbox{singlebox}{colback=green!10}

\begin{document}


%
\catchline{01}{01}{2003}{}{}
%

\title{Large Language Models for Fault Localization: An Empirical Study}

\author{Yingjian Xiao}
\address{School of Computer and Information Engineering, Nanchang Institute of Technology; Nanchang Key Laboratory of AI-Based Non-Destructive Phenotyping Measurement Technology and Equipment\\
Nanchang 330044, Jiangxi, China\\
\email{yingjianxiao@126.com}
}

\author{Weiwei Gong}
\address{School of Computer and Information Engineering,
Nanchang Institute of Technology; 
Nanchang Key Laboratory of AI-Based Non-Destructive Phenotyping Measurement Technology and Equipment\\
Nanchang 330044, Jiangxi, China\\
\email{51692825@qq.com}
}

\author{Jianjun Huang}
\address{School of Computer and Information Engineering,
Nanchang Institute of Technology; 
Nanchang Key Laboratory of AI-Based Non-Destructive Phenotyping Measurement Technology and Equipment\\
Nanchang 330044, Jiangxi, China\\
\email{27458808@qq.com}
}

\author{Rongqun Hu}
\address{School of Computer and Information Engineering,
Nanchang Institute of Technology; 
Nanchang Key Laboratory of AI-Based Non-Destructive Phenotyping Measurement Technology and Equipment\\
Nanchang 330044, Jiangxi, China\\
\email{25187632@qq.com}
}

\author{Hongwei Li}
\address{School of Artificial Intelligence,
Jiangxi Normal University\\
Nanchang 330022, Jiangxi, China\\
\email{lihongwei@jxnu.edu.cn}
}

\author{Anquan Jie}
\address{School of Artificial Intelligence,
Jiangxi Normal University\\
Nanchang 330022, Jiangxi, China\\
\email{janquan@163.com}
}

\author{Boyang Yang\footnote{Corresponding author.}}
\address{School of Artificial Intelligence (School of Software),
Yanshan University\\
Qinhuangdao 066004, Hebei, China\\
\email{yby@ieee.org}
}

\maketitle

\begin{history}
\end{history}

\begin{abstract}
Large Language Models (LLMs) have demonstrated strong performance on code-related tasks, particularly in automated program repair. However, repair effectiveness often depends on accurate upstream fault localization, while the statement-level fault localization capability of LLMs remains insufficiently evaluated. This paper presents a systematic empirical study of LLMs for statement-level fault localization. We evaluate four representative LLMs, including two open-weight models, Qwen2.5-Coder-32B-Instruct and DeepSeek-V3, and two closed-source models, GPT-4.1 mini and Gemini-2.5-Flash, on HumanEval-Java and Defects4J. The evaluation covers different input contexts and prompt strategies, including Zero-shot, Few-shot, and Chain-of-Thought prompting. We further assess model performance from three complementary perspectives: Exact Match, Partial Match, and output consistency, and compare LLMs with representative non-LLM baselines, including PMD and LineDef, under the same source-code-only input setting. In addition, we analyze practical efficiency and cost in terms of end-to-end response time and token-based API cost. The results show that bug report context improves observed fault localization performance on Defects4J; Few-shot prompting improves performance in some cases but does not yield consistent gains; and Chain-of-Thought prompting shows mixed effects across models. Overall, this study reveals the strengths, limitations, and practical trade-offs of LLMs in statement-level fault localization, providing empirical evidence for model selection and application in software engineering practice.
\end{abstract}


\input{1.intro}
\input{2.related}
\input{3.study_design}
\input{4.experimental_results}
\input{5.threats_to_validity}
\input{6.conclusions}

\section*{Acknowledgements}

This work was supported by the Science and Technology Research Project of Jiangxi Provincial Department of Education under Grant No. GJJ2402605.

\appendix
\input{7.Appendix}
\clearpage
\bibliography{sample}

\end{document}

%% file: 1.intro.tex
\section{Introduction}
\label{sec1}

With the rapid development of Large Language Models (LLMs) in the field of Natural Language Processing (NLP), their applications in software engineering have grown significantly, progressively expanding into various tasks such as code completion, unit test generation, and code summarization \cite{xuSystematicEvaluationLarge2022,chenEvaluatingLargeLanguage2021a,duEvaluatingLargeLanguage2024,haldarAnalyzingPerformanceLarge2024}. These models, through joint training on vast amounts of natural language and programming language data, have demonstrated robust capabilities in code understanding and generation, establishing a foundation for the development of intelligent and automated software engineering tools \cite{jiangSurveyLargeLanguage2024,bistarelliUsageLargeLanguage2025}.

Automated program repair (APR) has become a research hotspot in software engineering~\cite{yang2025survey}. Studies showed that LLMs can autonomously generate patches without fine-tuning, effectively repairing certain real faults \cite{yang2024cref,jiangEvaluatingFaultLocalization2024a,binmurtazaLLMFaultLocalisation2024a,luo2025unlocking}. For instance, Wu et al. \cite{wuLargeLanguageModels2023a} assessed ChatGPT’s repair capabilities, while Li et al. \cite{liEvaluatingGeneralizabilityLLMs2025a} tested 11 LLMs, including CodeLlama, on the Defects4J and DEFECTS4J-TRANS datasets. Additionally, a range of innovative methods have emerged, such as GiantRepair \cite{liHybridAutomatedProgram2025}, which integrates LLMs with program analysis to achieve efficient program repair through patch skeletons and context instantiation; CodeCorrector \cite{liContextawarePromptingLLMbased2025}, which infers repair directions based on test failure information and adaptively selects relevant code contexts to construct precise prompts for guiding patch generation; and ThinkRepair \cite{yinThinkRepairSelfDirectedAutomated2024a}, which implements automated program repair in two stages: first, by collecting repair examples with reasoning to build a knowledge base, and second, using Few-shot Learning (FSL) and iterative testing feedback to guide LLMs in generating correct patches. However, these studies primarily focus on the post-stage task of ``generating repairs'', while research on the pre-stage task of fault localization remains limited.

The goal of fault localization is to help developers quickly identify potential fault locations within code, making it a crucial step in the automated program repair process \cite{fanAutomatedRepairPrograms2023,louCanAutomatedProgram2020a,assiriFaultLocalizationAutomated2017,yang2025enhancing}. However, the application of LLMs in fault localization is still in the early stages of exploration. Only a few studies have provided preliminary evidence of models such as ChatGPT and CodeLlama’s potential in fault localization \cite{wuLargeLanguageModels2023a,liuEmpiricalEvaluationLarge2024,wangAssessingEffectivenessRecent2025,qinAgentFLScalingLLMbased2024}. However, these studies often lack a systematic and multi-perspective evaluation of LLM-based statement-level fault localization, particularly in terms of model comparison, prompt strategy analysis, and practical application costs. As such, the overall capability, applicability, and cost-effectiveness of LLMs in fault localization tasks remain significant research gaps.

To address these gaps, this paper presents a systematic empirical study of LLMs for statement-level fault localization. Specifically, we formulate the task as identifying buggy lines from a given Java source file under different input contexts and prompting strategies. We evaluate four representative LLMs, including GPT-4.1 mini, Qwen2.5-Coder-32B-Instruct, Gemini-2.5-Flash, and DeepSeek-V3, on HumanEval-Java and Defects4J under Zero-shot, Few-shot, and Chain-of-Thought prompting settings. Beyond exact localization success, we further examine partial-match performance, output consistency across repeated API calls, end-to-end response time, and token-based API cost. This multi-dimensional evaluation allows us to distinguish complete buggy-line localization from partial buggy-line identification, thereby providing a more nuanced understanding of the capability and limitations of LLMs in statement-level fault localization. All source code and experimental data are publicly available at {\url{https://github.com/XiaoYingJian/Empirical-LLM-Fault-Localization}}.

In summary, the main contributions of our study can be summarized as follows:

\begin{arabiclist}
\item We conduct a systematic empirical study of LLM-based statement-level fault localization across four representative LLMs, two benchmarks, multiple prompting strategies, different input contexts, and varying buggy-line counts. The study clarifies how these factors affect the localization capability of LLMs.

\item We provide a multi-perspective evaluation by jointly considering Exact Match (EM), Partial Match (PM), and output consistency measured by Consistency Rate (CR). The results highlight the distinction between complete buggy-line localization and partial buggy-line identification, showing that LLMs can still identify some ground-truth buggy lines in many cases even when exact localization remains limited.

\item We quantify the practical overhead of LLM-based fault localization in terms of end-to-end response time and token-based API cost. The results reveal trade-offs among localization performance, runtime efficiency, and API usage cost, supporting more informed model selection.
\end{arabiclist}

\textbf{Structure of the paper.} The remainder of this paper is organized as follows: Sec.~2 reviews related work on fault localization; Sec.~3 introduces the study design, including research questions, evaluation benchmarks, evaluation models and baselines, experimental settings, and evaluation metrics; Sec.~4 presents the experimental results and answers the research questions. Sec.~5 discusses Threats to Internal Validity, Threats to Construct Validity, and Threats to External Validity; Sec.~6 concludes the paper and suggests future research directions.

%% file: 2.related.tex
\section{Related Work}
\label{sec2}
This section reviews the literature and research progress related to this study in the following four areas: (1) traditional fault localization methods; (2) deep learning-based fault localization methods; (3) applications of Large Language Models in fault localization and program repair; (4) developments in prompt engineering and Chain-of-Thought reasoning in software engineering.

\subsection{Traditional Fault Localization Methods}
Traditional fault localization methods can generally be divided into static-analysis-based and dynamic-analysis-based approaches. Static analysis identifies potentially faulty or risky code locations by examining program structures and code properties without executing the program, such as syntax patterns, control-flow graphs (CFGs), data-flow relations, and rule violations \cite{wongSurveySoftwareFault2016}. In practice, many static analysis tools have been widely used to detect suspicious code patterns and report warning lines. For example, PMD detects potential issues in Java programs according to predefined rules and reports warnings with different priority levels \cite{trautschLongitudinalStudyStatic2020}. Amankwah et al. \cite{amankwahBugDetectionJava2023} conducted an extensive evaluation of eight static analysis and related bug-detection tools for Java programs, including FindBugs, PMD, YASCA, LAPSE+, JLint, Bandera, ESC/Java, and Java Pathfinder, and showed that different tools exhibit substantially different precision, recall, and false-positive characteristics. Dynamic-analysis-based approaches, in contrast, rely on runtime information such as test execution results and coverage profiles. A widely studied family of dynamic methods is Spectrum-Based Fault Localization (SBFL), which assigns suspiciousness scores to program elements, such as statements or methods, according to their execution frequencies in passing and failing test cases \cite{kimPreciseLearntoRankFault2019,pearsonEvaluatingImprovingFault2017}. Subsequent studies have further examined factors affecting SBFL effectiveness. For example, Sasaki et al. \cite{sasakiSBFLSuitabilitySoftwareCharacteristic2020} introduced SBFL-Suitability to characterize how program structure influences spectrum-based techniques, while Savant combines suspiciousness scores with program invariants and learning-to-rank strategies to rank potentially faulty methods \cite{b.leLearningtorankBasedFault2016}.

Overall, traditional fault localization methods are often efficient and interpretable for narrowing the inspection scope. However, rule-based static methods are constrained by predefined patterns, while dynamic methods depend on test coverage or runtime information. These characteristics may limit their ability to capture deeper semantic and task-specific debugging clues in complex or cross-module faults.

\subsection{Deep Learning-Based Fault Localization Methods}
In recent years, deep learning techniques have been widely applied to fault localization tasks, achieving better localization by learning code structure and semantic representations from large-scale labeled data. Yin et al. \cite{yinMultigraphLearningbasedSoftware2024a} proposed a multi-graph learning-based fault localization method, MGSDL, which represents code structure as a program dependency graph and combines it with semantic features extracted by CNNs to localize faults. Majd et al. \cite{majdSLDeepStatementlevelSoftware2020a} developed the SLDeep model, which utilizes 32 statement-level static features along with Long Short-Term Memory (LSTM) networks for fault prediction. The DeepLineDP model uses bidirectional GRU and hierarchical attention mechanisms for fault localization from the file level to the line level \cite{pornprasitDeepLineDPDeepLearning2023}; Li et al. \cite{liFaultLocalizationCode2021a} proposed DEEPRL4FL, which treats fault localization as an image pattern recognition problem, combining innovative code coverage representation learning and statement dependency analysis to locate faulty statements and methods accurately.

Although deep learning methods have achieved notable performance improvements, they rely heavily on large amounts of labeled data and generalize poorly to unseen data, making them difficult to apply directly to data-scarce or complex industrial systems.

\subsection{Applications of Large Language Models in Fault Localization and Program Repair}
In recent years, LLMs have demonstrated strong capabilities in code generation and program repair tasks. Wu et al. compared the performance of ChatGPT with traditional fault localization methods (e.g., SBFL, MBFL, and SmartFL) \cite{wuLargeLanguageModels2023a}; Yang et al. proposed the LLMAO framework, which fine-tunes a lightweight bidirectional adapter on a pre-trained LLM to perform fault localization without requiring test cases \cite{yangLargeLanguageModels2024a}; Kang et al. developed an automated fault localization method, AutoFL, which can accurately locate potential faults and generate natural language descriptions explaining "how the error occurred," providing a reasoning process that traditional methods lack \cite{kangQuantitativeQualitativeEvaluation2024a}; Jiang et al. evaluated several mainstream closed-source models, including ChatGPT, for fault localization and automated program repair tasks \cite{jiangEvaluatingFaultLocalization2024a}. Additionally, Xia et al. proposed ChatRepair, a conversational program repair method that uses ChatGPT for interactive patch generation and evaluates its performance and cost-effectiveness on datasets like Defects4J \cite{xiaAutomatedProgramRepair2024}.

While existing studies have shown the potential of LLMs in fault localization, there remains a lack of systematic empirical analysis of how different model access modes and prompt strategies affect fault localization performance. In addition, few studies have examined practical issues such as end-to-end response time and token-based API cost.

\subsection{Development of Prompt Engineering and Chain-of-Thought Reasoning in Software Engineering}
Prompt engineering has become a key technique for enhancing the performance of LLMs. Few-shot learning improves the model’s understanding of task structure and intent by providing a small number of example inputs \cite{brownLanguageModelsAre2020,wangGeneralizingFewExamples2021}; Chain-of-Thought reasoning guides the model through a step-by-step logical analysis, improving its ability to reason and explain complex tasks \cite{yangChainofThoughtNeuralCode2024,weiChainofThoughtPromptingElicits2022}. Zheng et al. proposed Few-VulD, a Few-shot learning-based framework for software vulnerability detection, addressing the problem of labeled data scarcity in software vulnerability detection \cite{zhengFewVulDFewshotLearning2024}; Wei et al. first proposed the CoT prompting strategy and achieved significant results in mathematical reasoning tasks \cite{weiChainofThoughtPromptingElicits2022}. Yang et al. \cite{yangChainofThoughtNeuralCode2024} introduced CoT into code generation tasks, fine-tuning lightweight models such as CodeLlama-7B to automatically generate high-quality code generation steps without requiring parameter updates to the models themselves; Rukmono et al. used CoT to guide LLMs through step-by-step reasoning, classifying code units into predefined reference architecture components for deductive software architecture recovery \cite{rukmonoDeductiveSoftwareArchitecture2024}.

Incorporating prompt engineering and CoT strategies into fault localization tasks has the potential to help LLMs capture task-specific debugging clues and produce more interpretable localization processes. Nevertheless, their effectiveness is not guaranteed and may depend on model characteristics, input context, dataset complexity, and fault types. This motivates a systematic empirical investigation of prompt-based LLM fault localization under different experimental settings.

%% file: 3.study_design.tex
\section{Study Design}\label{sec3}

\subsection{Research Questions}

To thoroughly investigate the capabilities and applicability of Large Language Models in fault localization tasks, this study poses the following four research questions:

\textbf{RQ1: What is the overall fault localization performance of different models in statement-level fault localization?} This study compares four representative LLMs with two categories of non-LLM baselines under the shared source-code-only setting. It further analyzes the effect of bug report context for the LLMs on Defects4J.

\textbf{RQ2: How does Few-shot learning affect the fault localization performance of LLMs?} This study examines the performance of LLMs under different Few-shot settings, namely One-shot, Two-shot, and Three-shot, and analyzes the impact of Few-shot learning on statement-level fault localization \cite{hannanDifficultySelectingFewShot2026,moreAnalysisLLMFineTuning2025}.

\textbf{RQ3: How does Chain-of-Thought prompting affect the fault localization performance of LLMs?} This study analyzes the changes in model localization results after introducing the CoT strategy, and investigates the impact of explicit step-by-step reasoning guidance on statement-level fault localization performance \cite{yangChainofThoughtNeuralCode2024}.

\textbf{RQ4: What are the end-to-end response time and token-based API costs of LLMs in fault localization?} This study measures and compares end-to-end response time under different prompt strategies and analyzes token-based API cost under the Zero-shot source-code-only setting, with the aim of assessing their runtime efficiency and API usage cost.

\subsection{Evaluation Benchmarks}
This study uses HumanEval-Java and Defects4J V1.2.0 as evaluation benchmarks to assess the statement-level fault localization performance of LLMs. Table~1 summarizes the basic information of the two datasets.

The HumanEval-Java dataset contains 163 Java bugs, ranging from simple operator errors to complex logic faults that require multi-line modifications. These bugs are manually injected into correct Java programs and are accompanied by corresponding JUnit test cases. Owing to its diverse bug types and well-defined task formulation, we use HumanEval-Java as a controlled, manually constructed Java bug benchmark to evaluate model performance in statement-level fault localization \cite{jiangImpactCodeLanguage2023a,vallecillosruizAssessingLatentAutomated2025,huang2025template}.

Defects4J V1.2.0 \cite{justDefects4JDatabaseExisting2014a} is an industrial-grade open-source Java benchmark widely used in software fault localization and repair research. It contains 395 real faults from six open-source projects, including Chart, Closure, and Lang. Each fault is accompanied by a complete bug report, reproducible test cases, and the corresponding fixed version, providing a standardized benchmark for evaluating automated fault localization and repair techniques. Due to these advantages, Defects4J has been widely adopted in software engineering research \cite{qinSoapFLStandardOperating2025,zhangContextbasedTransferLearning2025,louCanAutomatedProgram2020a,liDeepFLIntegratingMultiple2019}. By conducting fault localization experiments on Defects4J, this study examines the performance and generalization ability of LLMs in more realistic software engineering scenarios, with the aim of evaluating their practical effectiveness and adaptability.

\begin{table}[h]
\tbl{Overview of the benchmark datasets used in this study}
{
\begin{tabular}{@{}cccc@{}}
\toprule
Dataset Name & Number of Bugs & Source &Year \\
\colrule
HumanEval-Java & 163 & Manually Generated & 2022 \\
Defects4J V1.2.0 & 395 & Real Projects & 2018 \\
\botrule
\end{tabular}
}
\label{tab:datasets}
\end{table}

\subsection{Evaluation Models and Baselines}
To systematically evaluate the performance of Large Language Models (LLMs) in statement-level fault localization, this study selects four representative LLMs for evaluation, including two closed-source models, GPT-4.1 mini and Gemini-2.5-Flash, and two open-weight models, Qwen2.5-Coder-32B-Instruct and DeepSeek-V3. These models have demonstrated strong capabilities in code understanding, code generation, program repair, and reasoning tasks, thereby enabling comparisons of LLM-based fault localization across different model categories and serving modes. Table 2 summarizes the basic information of the four LLMs.

\begin{itemize}
\item[\textendash] \textbf{GPT-4.1 mini.} GPT-4.1 mini is a lightweight model released by OpenAI. It is optimized for response speed and resource efficiency, making it suitable for software engineering tasks such as code analysis, test generation, and program repair \cite{openai2025gpt41,openai2026gpt41mini}.

\item[\textendash] \textbf{Qwen2.5-Coder-32B-Instruct.} Qwen2.5-Coder-32B-Instruct is released by Alibaba and optimized for code understanding and generation tasks. It contains approximately 32.5 billion parameters, is built on the Qwen2.5 backbone, and is further instruction-tuned. The model supports multilingual programming tasks and performs well on several code-related benchmarks \cite{xuKodCodeDiverseChallenging2025}.

\item[\textendash] \textbf{Gemini-2.5-Flash.} Gemini-2.5-Flash is released by Google DeepMind. It supports multimodal inputs, including text, code, and images, and is designed to provide efficient reasoning and low-latency responses. These characteristics make it suitable for software engineering tasks such as complex code analysis and fault localization \cite{jukiewiczSentimentAnalysisLarge2025,guoSWEFactoryYourAutomated2025}.

\item[\textendash] \textbf{DeepSeek-V3.} DeepSeek-V3 is released by the DeepSeek team and adopts a mixed pre-training strategy over natural language and programming language corpora. It supports both Chinese and English inputs and demonstrates strong code understanding and generation capabilities, making it applicable to a wide range of software engineering tasks \cite{dengExploringDeepSeekSurvey2025,silvaRepairBenchLeaderboardFrontier2025}.
\end{itemize}
To compare LLM-based fault localization with representative non-LLM methods under the source-code-only setting, this study selects PMD and LineDef as baselines, corresponding to traditional static analysis and learning-based line-level fault localization, respectively.

\begin{itemize}
\item[\textendash] \textbf{PMD.} PMD \cite{trautschLongitudinalStudyStatic2020,guoCodelinelevelBugginessIdentification2023} is a rule-based Java static analysis tool that detects potential issues in source code according to predefined rules and reports warning line numbers with severity levels. In this study, PMD is used as a traditional static analysis baseline to evaluate the performance of rule-driven methods in statement-level fault localization.

\item[\textendash] \textbf{LineDef.} LineDef \cite{yinLinelevelDefectPrediction2025} is a graph convolutional network-based line-level defect prediction method. It captures contextual semantic relationships in code by constructing token-level and line-level code graphs, and incorporates an attention mechanism to generate risk scores for code lines. Since these risk scores can be converted into line-level suspiciousness rankings, we adapt LineDef as a learning-based ranking baseline for statement-level fault localization.
\end{itemize}

\begin{table}[h]
\tbl{Overview of the LLMs evaluated in this study}
{
\begin{tabular}{@{}ccccc@{}}
\toprule
Model & Model Size & Release Time & Knowledge Cutoff & Model Access \\
\colrule
GPT-4.1 mini & Not Disclosed & 2025-04 & 2024-06 & Closed-source \\
Qwen2.5-Coder-32B-Instruct & 32.5B & 2024-11 & 2023-10 & Open-weight \\
Gemini-2.5-Flash & Not Disclosed & 2025-04 & 2025-01 & Closed-source \\
DeepSeek-V3 & 671B & 2024-12 & 2024-07 & Open-weight \\
\botrule
\end{tabular}
}
\label{tab:models}
\end{table}

\subsection{Experimental Settings}
To improve the reproducibility of API-based experiments, this study uses standardized task instructions and setting-specific user-prompt templates for all LLM experiments. The complete Zero-shot, Few-shot, and CoT prompt templates are provided in Appendix B. The temperature parameter of all models is set to 0 to reduce decoding randomness and to compare the evaluated models under consistent reported API settings. However, temperature = 0 is not treated as a guarantee of fully deterministic outputs in API-based LLM services. As shown in Appendix~A, non-identical localized buggy-line sets are still observed across repeated API calls under the same prompt and the same reported API settings. Unless otherwise specified, other decoding-related parameters, such as \texttt{top\_p} and \texttt{max\_tokens}, are kept at the default values provided by the corresponding API provider. For each API call, we record the raw model response, prompt tokens, completion tokens, start timestamp, end timestamp, and end-to-end response time. The raw model output is then parsed and normalized using a unified post-processing procedure, and the resulting localized buggy-line set is mapped back to the corresponding source-code lines for subsequent calculation of evaluation metrics. The detailed parsing and exact-match rules are described in Section~3.5.1.

For repeated-call experiments, each bug case is queried 13 times under the same prompt and the same reported API settings. These repeated calls are used to characterize the empirical success probability and output consistency of API-based LLM fault localization under a fixed API configuration, rather than to evaluate output diversity under high-temperature sampling. The results of these repeated calls are used to compute the exact-match-based metrics and output consistency metrics defined in Section~3.5. The detailed prompt design, dataset-specific input settings, and baseline evaluation protocols are described in the corresponding experimental design sections and Appendix~B.

\subsection{Evaluation Metrics}
This study adopts two categories of evaluation metrics: Exact Match and Partial Match. EM-based metrics assess whether a model can identify the complete set of ground-truth buggy lines, whereas PM-based metrics further quantify the overlap between localized candidate lines and ground-truth buggy lines.
\subsubsection{Exact Match Metrics}
In the LLM-based repair scenario, the completeness of fault localization results directly affects the efficiency of subsequent patch generation. If key buggy lines are missed, it may lead to incomplete repairs, repeated model calls, or additional manual inspection. Therefore, this study adopts EM-based metrics to evaluate the ability of models to identify the complete buggy-line set. For each bug case, a localization result is considered successful only when the predicted buggy-line set exactly matches the ground-truth buggy-line set. In implementation, model outputs are parsed according to the required format, i.e., ``Line X: bug statement'', and minor formatting differences such as extra spaces, quotation marks, and Markdown markers are normalized before matching. The predicted line numbers are then mapped back to the original source-code lines. Outputs with invalid line numbers or outputs that cannot be mapped to source-code lines are discarded. An EM success is counted only when the normalized predicted line-number set is identical to the ground-truth buggy-line set. Based on the 13 repeated calls described in Section~3.4, this study uses Top@k, Pass@k, and CR to evaluate exact-match success and output consistency.
\begin{itemize}

\item[\textendash] \textbf{Top@k:} \(\mathrm{Top@}k\) measures whether the model produces at least one localization result that exactly matches all ground-truth buggy lines within the first \(k\) repeated calls. This study reports \(\mathrm{Top@}5\) and \(\mathrm{Top@}10\) to reflect exact-match success under a limited number of actual API calls~\cite{luoWhenFineTuningLLMs2025a}.

\item[\textendash] \textbf{Pass@k:} Pass@k estimates the probability that the model obtains at least one correct localization result within $k$ attempts, based on $n$ repeated calls. In this study, Pass@k is used as an empirical repeated-call metric under a fixed API configuration. Unlike its common use in sampling-based code generation evaluation, our calculation is based on repeated API calls made with the same prompt and the same API configuration. It therefore estimates whether at least one exact-match localization result can be observed within $k$ attempts. It is computed as follows:
\begin{equation}
\mathrm{Pass@}k =
\mathbb{E}_{\mathrm{problems}}
\left[
1 - \frac{\binom{n-c}{k}}{\binom{n}{k}}
\right]
\label{eq:pass_at_k}
\end{equation}

where \(n\) denotes the total number of localization results generated for each bug case. In this study, \(n=13\). \(c\) denotes the number of localization results that satisfy the EM criterion, and \(k\) denotes the number of attempts considered. Following Luo et al.~\cite{luoWhenFineTuningLLMs2025a}, we set \(k=1,5,10\). Kochhar et al. reported that developers rarely inspect more than five candidate patches~\cite{kochharPractitionersExpectationsAutomated2016}, while Noller et al. found that the practical upper bound of patches developers are willing to review is approximately ten~\cite{nollerTrustEnhancementIssues2022}.

\item[\textendash] \textbf{Consistency Rate (CR):} CR measures the consistency of localized buggy-line sets for the same bug case across repeated calls. Inspired by the definition of intra-prompt stability proposed by Carandang et al.~\cite{carandangAreLLMsReliable2025}, we adapt CR to the fault localization setting. Specifically, for each bug case, we first compute \(CR_h\), which denotes the proportion of all pairwise combinations of repeated calls whose localized buggy-line sets are exactly identical for the \(h\)-th bug case:

\begin{equation}
CR_h =
\frac{
\sum_{i,j \in \binom{k}{2}} \mathbf{1}(P_{h,i}=P_{h,j})
}{
\binom{k}{2}
}
\times 100
\end{equation}

where \(k\) denotes the number of repeated calls used for the CR calculation. In this study, \(k=13\). \(h\in[1,N]\), where \(N\) denotes the number of bug cases. \(P_{h,i}\) denotes the localized buggy-line set produced by the \(i\)-th call for the \(h\)-th bug case, and \(\mathbf{1}(\cdot)\) is an indicator function that equals 1 if the condition inside the parentheses holds and 0 otherwise. We report the median and interquartile range of all \(CR_h\) values, denoted as \(CR_{\mathrm{median}}\pm IQR\), to characterize the central tendency and dispersion of output consistency.

\(\mathrm{Pass@}k\) and CR capture different aspects of repeated API calls. \(\mathrm{Pass@}k\) measures whether repeated invocations can obtain at least one exact-match localization result, whereas CR measures whether the localized buggy-line sets remain identical across repeated invocations.

Unless otherwise specified, all LLM evaluation results are computed based on 13 repeated calls. Precision, Recall, and F1-score are reported as mean \(\pm\) standard deviation. PMD is a deterministic static analysis tool, and therefore only single-run results are reported. LineDef is trained and tested 13 times, and its results are reported as mean \(\pm\) standard deviation.

\end{itemize}

\subsubsection{Partial Match Metrics}

In practical debugging, developers often use fault localization results to narrow the code inspection scope and prioritize code lines that are more likely to contain faults. For multi-line faults in particular, even if a model identifies only a subset of key buggy lines, the localization result may still provide useful clues for manual inspection, fault understanding, and subsequent patch generation. Therefore, this study further introduces line-level Precision, Recall, and F1-score as PM-based metrics to evaluate the overlap between localized candidate buggy lines and ground-truth buggy lines, thereby complementing EM-based metrics and reflecting the practical utility of partially correct localization results.

For each bug case, let \(P\) denote the localized buggy-line set and \(G\) denote the ground-truth buggy-line set. Precision, Recall, and F1-score are defined as follows:

\begin{equation}
\mathrm{Precision}=\frac{|P\cap G|}{|P|}
\end{equation}

\begin{equation}
\mathrm{Recall}=\frac{|P\cap G|}{|G|}
\end{equation}

\begin{equation}
\mathrm{F1}=\frac{2\times \mathrm{Precision}\times \mathrm{Recall}}{\mathrm{Precision}+\mathrm{Recall}}
\end{equation}

For each evaluated bug case, Precision, Recall, and F1-score are first computed separately according to Eqs. (3)--(5). The reported Precision, Recall, and F1-score are then obtained by averaging the corresponding per-case values over all evaluated bug cases. In other words, F1-score is averaged as an individual metric, rather than being recalculated from the averaged Precision and Recall.

%% file: 4.experimental_results.tex
\section{Experimental Results and Analysis}
\label{sec4}
\subsection{RQ1: Overall Fault Localization Performance}
\textbf{[Experiment Goal]}: This experiment aims to evaluate the overall performance of different types of methods in statement-level fault localization. Specifically, we compare four representative LLMs with two categories of non-LLM baselines under the shared source-code-only setting on the HumanEval-Java and Defects4J datasets. We further analyze the effect of bug report context for LLMs on Defects4J and examine how the number of ground-truth buggy lines affects localization performance. Through this experiment, we seek to characterize the empirical strengths and limitations of LLM-based fault localization relative to traditional static analysis methods and deep learning-based line-level fault localization methods, while distinguishing direct baseline comparisons from LLM-specific contextual analysis.

\textbf{[Experiment Design]}: We use the Zero-shot prompting strategy for a unified evaluation. Since HumanEval-Java does not provide bug report information, all experiments on this dataset use only source code as input. In contrast, Defects4J provides bug report information; therefore, we consider two input settings, without and with bug report context, to analyze the effect of contextual information. The complete Zero-shot prompt template is provided in Appendix~B. For each bug case, we perform 13 independent API calls under the same prompt and API parameter settings, and compute model performance according to the metrics defined in Section~3.5.

For non-LLM baselines, we convert the outputs of PMD and LineDef into line-level suspiciousness rankings. PMD produces warning lines according to static rules and ranks candidate buggy lines based on warning priority. Since its output is deterministic, we report only a single-run result for PMD. LineDef is evaluated under a cross-project prediction setting, where the original project data provided by its authors are used for training, and HumanEval-Java and Defects4J are used as external test sets. To reduce the impact of training randomness, we repeat the training and testing process 13 times and report the mean $\pm$ standard deviation. For both baselines, we adopt an offline Top-$m$ protocol to convert their ranked suspiciousness lists into localized candidate-line sets, where $m$ equals the number of ground-truth buggy lines in each defective file. This setting controls the candidate-set size for partial-match evaluation and is not intended as a deployable configuration. Precision, Recall, and F1-score are then computed based on the selected lines. It should be noted that direct comparison with PMD and LineDef is conducted only under the shared source-code-only setting, because PMD and LineDef operate on source code and are not designed to consume natural-language bug report context. Therefore, the Defects4J results obtained with bug report context are reported as an LLM-specific contextual analysis and should not be directly compared with PMD or LineDef.

In addition, to analyze the effect of buggy-line count on LLM localization performance, we group localization samples according to the number of ground-truth buggy lines in each defective Java file. Specifically, the samples are divided into four groups: 1-line, 2-line, 3-line, and $\geq$4-line defective files, and Pass@10 is reported for each group. Since some Defects4J bugs involve multiple modified Java files, the number of defective-file samples is larger than the number of Defects4J bug IDs.

\textbf{[Experimental Results]:} As shown in Table~3, under the Zero-shot setting, the four LLMs achieve relatively high EM-based performance on HumanEval-Java, but localization performance and output consistency do not necessarily align. Gemini-2.5-Flash achieves the highest observed EM-based performance among the evaluated models under this setting, with Top@5, Top@10, and Pass@10 reaching 65.03\%, 67.48\%, and 68.73\%, respectively. However, its \(CR_{\mathrm{median}}\) is 71.79\%, which is lower than those of GPT-4.1 mini and DeepSeek-V3. In contrast, GPT-4.1 mini and DeepSeek-V3 both achieve a \(CR_{\mathrm{median}}\) of 100.00\%, suggesting that their outputs are more stable for typical bug cases. However, their Pass@10 values are 51.37\% and 54.69\%, respectively, which are lower than that of Gemini-2.5-Flash. These results indicate that, on HumanEval-Java, higher output consistency does not necessarily imply a higher localization success rate. A model may consistently produce the same prediction, but that prediction may not be the correct localization result.

On Defects4J, the EM-based performance of all four models is substantially lower than that on HumanEval-Java, indicating that fault localization in real-world projects is more challenging. Without bug report context, the highest Pass@10 among all models is only 5.83\%, and \(CR_{\mathrm{median}}\) is generally low. In particular, Qwen2.5-Coder-32B-Instruct and DeepSeek-V3 achieve \(CR_{\mathrm{median}}\) values of 38.46\% and 35.90\%, respectively, suggesting that without contextual information, the models not only struggle to identify the complete set of buggy lines but also produce less stable outputs. After bug report context is introduced, all models show improved localization performance. Gemini-2.5-Flash achieves Top@5 and Pass@10 values of 23.78\% and 23.92\%, respectively. Meanwhile, the \(CR_{\mathrm{median}}\) values of all models also increase. Nevertheless, the IQR values of most models remain large, indicating that bug report context improves output consistency for typical cases, while stability still varies substantially across different bug cases.

As shown in Table~4, under the shared source-code-only setting, the evaluated LLMs show higher observed partial-match scores than PMD and LineDef. On HumanEval-Java, the F1-scores of the four LLMs are close to or above 59\%, higher than those of LineDef and PMD. On Defects4J without bug report context, the Precision, Recall, and F1-score of all models decrease substantially, but the evaluated LLMs still show higher observed partial-match performance than the two non-LLM baselines under the same source-code-only condition. The results with bug report context are reported only for the LLMs, because PMD and LineDef do not consume bug reports. After bug report context is added, the partial-match performance of LLMs further improves, with Gemini-2.5-Flash achieving the highest Precision and F1-score of 40.89\% and 31.87\%, respectively. These results suggest that, although LLMs remain limited in complete buggy-line localization, they can identify a subset of suspicious lines and may help reduce the manual inspection scope in practical debugging.

As shown in Table~5, the number of buggy lines in a defective file has a clear effect on localization performance. On HumanEval-Java, where each bug corresponds to one defective Java file, all models achieve relatively high Pass@10 values for 1-line defective files, with Gemini-2.5-Flash reaching 77.00\%. When the number of buggy lines increases to two or three, the performance of most models declines markedly. For defective files with $\geq$4 buggy lines, the Pass@10 of some models increases again; however, this group contains relatively few samples. On Defects4J, the performance degradation caused by increasing buggy-line count is more pronounced. When a defective file involves three or more buggy lines, Pass@10 drops sharply. Even with bug report context, the localization success rate for defective files with $\geq$4 buggy lines remains close to zero. These findings indicate that existing LLMs still have clear limitations in handling complex multi-line defective files in real-world projects.

\begin{table*}[t]
\centering
\setlength{\tabcolsep}{3pt}
\renewcommand{\arraystretch}{1.12}
\tbl{Performance Comparison under the Zero-shot Setting (\%)}
{
\resizebox{\textwidth}{!}{
\begin{tabular}{@{}ccccccccc@{}}
\toprule
Dataset & Input setting & Model & Top@5 & Top@10 & Pass@1 & Pass@5 & Pass@10 & CR$_{median}$ $\pm$ IQR \\
\colrule
\multirow{4}{*}{HumanEval-Java}
& \multirow{4}{*}{Source code only}
& GPT-4.1 mini & 50.31 & 50.92 & 45.97 & 50.77 & 51.37 & $100.00 \pm 33.97$ \\
& & Qwen2.5-Coder-32B-Instruct & 46.63 & 50.31 & 37.66 & 47.71 & 50.18 & $71.79 \pm 67.95$ \\
& & Gemini-2.5-Flash & 65.03 & 67.48 & 52.90 & 65.59 & 68.73 & $71.79 \pm 51.92$ \\
& & DeepSeek-V3 & 52.15 & 53.37 & 46.15 & 52.88 & 54.69 & $100.00 \pm 38.46$ \\
\midrule
\multirow{8}{*}{Defects4J}
& \multirow{4}{*}{Without Bug Report}
& GPT-4.1 mini & 4.20 & 4.20 & 3.16 & 4.11 & 4.36 & $53.85 \pm 73.08$ \\
& & Qwen2.5-Coder-32B-Instruct & 4.90 & 5.59 & 3.16 & 4.69 & 5.32 & $38.46 \pm 39.86$ \\
& & Gemini-2.5-Flash & 5.83 & 5.83 & 4.55 & 5.72 & 5.83 & $48.72 \pm 15.38$ \\
& & DeepSeek-V3 & 2.10 & 2.56 & 1.29 & 2.45 & 2.91 & $35.90 \pm 75.64$ \\
\cmidrule(lr){2-9}
& \multirow{4}{*}{With Bug Report}
& GPT-4.1 mini & 15.15 & 15.62 & 12.96 & 15.02 & 15.45 & $70.51 \pm 63.64$ \\
& & Qwen2.5-Coder-32B-Instruct & 13.75 & 15.38 & 9.58 & 13.78 & 15.01 & $53.85 \pm 56.41$ \\
& & Gemini-2.5-Flash & 23.78 & 24.48 & 17.89 & 22.60 & 23.92 & $57.69 \pm 62.82$ \\
& & DeepSeek-V3 & 11.66 & 11.66 & 9.61 & 11.06 & 11.56 & $71.79 \pm 61.83$ \\
\botrule
\end{tabular}
}
}
\label{tab:humaneval_zero}
\label{tab:defects4j_zero}
\end{table*}

\begin{table*}[t]
\centering
\setlength{\tabcolsep}{3pt}
\renewcommand{\arraystretch}{1.12}
\tbl{Partial Match Performance under the Zero-shot Setting (\%)}
{
\resizebox{\textwidth}{!}{
\begin{tabular}{@{}cccccccccc@{}}
\toprule
\multirow{3}{*}{Model}
& \multicolumn{3}{c}{HumanEval-Java}
& \multicolumn{6}{c}{Defects4J} \\
\cmidrule(lr){2-4} \cmidrule(lr){5-10}
& \multicolumn{3}{c}{}
& \multicolumn{3}{c}{Without Bug Report}
& \multicolumn{3}{c}{With Bug Report} \\
\cmidrule(lr){5-7} \cmidrule(lr){8-10}
& Precision & Recall & F1-score
& Precision & Recall & F1-score
& Precision & Recall & F1-score \\
\colrule
GPT-4.1 mini
& $70.56 \pm 0.85$ & $77.72 \pm 0.58$ & $70.34 \pm 0.63$
& $9.99 \pm 0.40$ & $12.83 \pm 0.83$ & $8.48 \pm 0.42$
& $28.80 \pm 0.82$ & $31.61 \pm 1.35$ & $24.68 \pm 0.76$ \\
Qwen2.5-Coder-32B-Instruct
& $63.59 \pm 1.26$ & $63.70 \pm 1.25$ & $59.26 \pm 0.95$
& $8.40 \pm 0.69$ & $8.21 \pm 0.93$ & $6.29 \pm 0.55$
& $20.49 \pm 0.99$ & $19.41 \pm 1.30$ & $16.46 \pm 0.88$ \\
Gemini-2.5-Flash
& $72.09 \pm 1.61$ & $71.38 \pm 1.17$ & $69.67 \pm 1.29$
& $13.43 \pm 0.59$ & $14.74 \pm 0.40$ & $11.32 \pm 0.43$
& $40.89 \pm 0.74$ & $32.59 \pm 0.92$ & $31.87 \pm 0.67$ \\
DeepSeek-V3
& $70.26 \pm 0.96$ & $76.17 \pm 0.94$ & $69.59 \pm 0.71$
& $6.50 \pm 0.53$ & $20.67 \pm 1.13$ & $5.28 \pm 0.49$
& $23.35 \pm 0.59$ & $25.43 \pm 0.55$ & $17.98 \pm 0.45$ \\
LineDef
& $12.54 \pm 3.26$ & $12.54 \pm 3.26$ & $12.54 \pm 3.26$
& $2.26 \pm 0.71$ & $2.26 \pm 0.71$ & $2.26 \pm 0.71$
& - & - & - \\
PMD
& 2.86 & 2.49 & 2.64
& 1.86 & 1.26 & 1.37
& - & - & - \\
\botrule
\end{tabular}
}
}
\label{tab:zero_partial_match}
\end{table*}

\begin{table*}[!htbp]
\centering
\setlength{\tabcolsep}{4pt}
\renewcommand{\arraystretch}{1.12}
\tbl{Pass@10 by Number of Buggy Lines in Defective Files under the Zero-shot Setting (\%)}
{
\resizebox{\textwidth}{!}{
\begin{tabular}{@{}cccccccc@{}}
\toprule
Dataset & \shortstack{Input\\setting} & \shortstack{Buggy lines/file} & Total count & \shortstack{GPT-4.1\\mini} & \shortstack{Qwen2.5-Coder-\\32B-Instruct} & \shortstack{Gemini-2.5-\\Flash} & DeepSeek-V3 \\
\colrule
\multirow{4}{*}{HumanEval-Java}
& \multirow{4}{*}{Source code only}
& 1 & 102 & 57.58 & 61.60 & 77.00 & 60.48 \\
& & 2 & 40 & 47.50 & 32.39 & 59.31 & 49.22 \\
& & 3 & 13 & 15.38 & 15.38 & 46.15 & 30.77 \\
& & \(\geq 4\) & 8 & 50.00 & 50.00 & 47.12 & 47.12 \\
\midrule
\multirow{8}{*}{Defects4J}
& \multirow{4}{*}{Without Bug Report}
& 1 & 198 & 8.08 & 10.24 & 10.10 & 4.31 \\
& & 2 & 83 & 3.29 & 3.06 & 4.82 & 3.56 \\
& & 3 & 42 & 0.00 & 0.00 & 2.38 & 2.37 \\
& & \(\geq 4\) & 106 & 0.00 & 0.00 & 0.00 & 0.00 \\
\cmidrule(lr){2-8}
& \multirow{4}{*}{With Bug Report}
& 1 & 198 & 27.91 & 28.22 & 46.11 & 19.50 \\
& & 2 & 83 & 10.84 & 8.39 & 11.21 & 12.05 \\
& & 3 & 42 & 4.76 & 0.00 & 4.76 & 0.00 \\
& & \(\geq 4\) & 106 & 0.00 & 1.45 & 0.00 & 0.94 \\
\botrule
\multicolumn{8}{@{}p{\textwidth}@{}}{\footnotesize\raggedright Note: Total count refers to defective Java files, not bug IDs.} \\
\end{tabular}
}
}
\label{tab:humaneval_bugline}
\label{tab:defects4j_bugline}
\end{table*}

\noindent
\fbox{%
\begin{minipage}{\dimexpr\textwidth-2\fboxsep-2\fboxrule}
\textbf{Finding 1}: LLMs remain limited in complete buggy-line localization, but they show stronger capability in partial buggy-line identification, which can provide useful suspicious-line clues and may help reduce the manual inspection scope in practical debugging.

\medskip

\textbf{Finding 2}: Bug report context improves observed fault localization performance in complex projects. The improvement is most evident for 1-line and 2-line defective files, and for some 3-line defective files, bug report context can increase Pass@10 from 0\% to non-zero values.

\medskip
\textbf{Finding 3}: The number of buggy lines in a defective file is generally associated with task difficulty. On HumanEval-Java, 3-line defective files lead to a clear performance drop across models. On Defects4J, Pass@10 remains in the single digits once defective files involve three or more buggy lines, and defective files with $\geq$4 buggy lines remain largely unresolved even with bug report context.

\medskip
\textbf{Finding 4}: Controlled decoding does not guarantee stable LLM-based fault localization outputs. Even under the same prompt and temperature setting, CR varies with task complexity and input context: output consistency is lower for complex project code without bug report context, whereas adding bug report context increases \(CR_{\mathrm{median}}\) for all evaluated models. This suggests that output stability is shaped not only by decoding parameters, but also by code complexity and contextual completeness.
\end{minipage}
}

\subsection{RQ2: Effect of Few-shot Learning}
\textbf{[Experiment Goal]}: This experiment aims to evaluate the effect of Few-shot learning on the statement-level fault localization performance of LLMs. By comparing the localization performance of four LLMs under the One-shot, Two-shot, and Three-shot settings, we analyze how the presence and number of in-context examples affect model localization results.

\textbf{[Experiment Design]}: We use the same four LLMs and construct One-shot, Two-shot, and Three-shot prompts for the HumanEval-Java and Defects4J datasets. For each bug case, we perform 13 repeated API calls under the same prompt and API parameter settings, and compute model performance according to the evaluation metrics defined in Section~3.5. Since HumanEval-Java does not provide bug report information, its Few-shot prompts consist only of source code and in-context examples. In contrast, Defects4J provides bug report information; therefore, its Few-shot prompts include source code, bug report context, and in-context examples. For the same target case, the in-context examples were kept identical across all models. These examples were selected from different bug cases and did not share the same bug ID or defective file with the target case. No ground-truth information from the target case was used in prompt construction. The complete Few-shot prompt templates are provided in Appendix~B.

\textbf{[Experimental Results]}: As shown in Table~6, the effect of Few-shot learning on EM-based performance and output consistency does not increase linearly with the number of examples. On HumanEval-Java, GPT-4.1 mini achieves its best performance under the Two-shot setting, with a Pass@10 of 56.58\%. Meanwhile, its \(CR_{\mathrm{median}}\) reaches 100.00\%, with an IQR of 0.00, indicating that this setting improves exact-match localization capability while maintaining stable outputs. DeepSeek-V3 achieves its highest Pass@10 under the Three-shot setting, reaching 58.58\%, with a \(CR_{\mathrm{median}}\) of 100.00\%. In contrast, although Qwen2.5-Coder-32B-Instruct achieves a \(CR_{\mathrm{median}}\) of 100.00\% under multiple Few-shot settings, its Pass@10 does not consistently improve, suggesting that high output consistency does not necessarily correspond to higher localization accuracy. Gemini-2.5-Flash maintains relatively high EM-based performance under Few-shot settings, but it does not yield consistent gains over the Zero-shot setting.

On Defects4J, because Few-shot prompts include bug report context, the results should mainly be compared with the Zero-shot setting that also includes bug report context. Qwen2.5-Coder-32B-Instruct achieves the highest Pass@10 under the Two-shot setting, reaching 17.61\%, but its \(CR_{\mathrm{median}}\) is only 39.74\%, indicating that the performance gain is accompanied by considerable output variability. Gemini-2.5-Flash maintains both relatively high Pass@10 and high \(CR_{\mathrm{median}}\) under the One-shot and Two-shot settings, showing a better balance between performance and stability. For GPT-4.1 mini and DeepSeek-V3, the Few-shot results show only limited improvement over the Zero-shot setting with bug report context, and their IQR values are generally large, indicating that output stability still varies across different bug cases. Overall, a small number of in-context examples may help models understand the task structure and expected output format, but adding more examples does not necessarily lead to a higher exact-match success rate.

As shown in Table~7, the effect of Few-shot learning on PM-based metrics is more stable than its effect on EM-based metrics. On HumanEval-Java, GPT-4.1 mini and Qwen2.5-Coder-32B-Instruct achieve their highest F1-scores under the Two-shot setting, reaching 72.28\% and 63.32\%, respectively. DeepSeek-V3 achieves its highest F1-score under the Three-shot setting, reaching 68.33\%. Gemini-2.5-Flash shows generally stable partial-match performance across Few-shot settings. On Defects4J, the Three-shot setting usually improves Recall and F1-score. The F1-scores of Qwen2.5-Coder-32B-Instruct, Gemini-2.5-Flash, and DeepSeek-V3 all improve compared with the One-shot setting. These results suggest that additional examples may help models cover more ground-truth buggy lines, but such partial-match gains do not necessarily translate into improvements in exact-match success.

\begin{table*}[!htbp]
\centering
\setlength{\tabcolsep}{3pt}
\renewcommand{\arraystretch}{1.08}
\tbl{Exact Match Performance under Few-shot Settings (\%)}
{
\resizebox{\textwidth}{!}{
\begin{tabular}{@{}ccccccccc@{}}
\toprule
Dataset & Model & Setting & Top@5 & Top@10 & Pass@1 & Pass@5 & Pass@10 & CR$_{median}$ $\pm$ IQR \\
\colrule
\multirow{12}{*}{HumanEval-Java}
& \multirow{3}{*}{GPT-4.1 mini}
& One-shot & 53.99 & 54.60 & 49.98 & 53.46 & 54.76 & $100.00 \pm 15.38$ \\
& & Two-shot & 54.60 & 56.44 & 51.91 & 55.24 & 56.58 & $100.00 \pm 0.00$ \\
& & Three-shot & 50.92 & 52.15 & 49.13 & 51.60 & 52.45 & $100.00 \pm 0.00$ \\
\cmidrule(lr){2-9}
& \multirow{3}{*}{Qwen2.5-Coder-32B-Instruct}
& One-shot & 49.08 & 50.92 & 42.80 & 49.09 & 50.92 & $100.00 \pm 33.97$ \\
& & Two-shot & 47.24 & 49.08 & 41.67 & 47.61 & 48.87 & $100.00 \pm 39.74$ \\
& & Three-shot & 46.63 & 49.08 & 41.86 & 47.28 & 48.74 & $100.00 \pm 15.38$ \\
\cmidrule(lr){2-9}
& \multirow{3}{*}{Gemini-2.5-Flash}
& One-shot & 65.03 & 68.71 & 49.74 & 63.89 & 68.27 & $84.62 \pm 53.85$ \\
& & Two-shot & 64.42 & 68.10 & 52.05 & 64.92 & 67.45 & $84.62 \pm 53.21$ \\
& & Three-shot & 63.80 & 66.87 & 50.59 & 63.60 & 67.78 & $84.62 \pm 53.85$ \\
\cmidrule(lr){2-9}
& \multirow{3}{*}{DeepSeek-V3}
& One-shot & 51.53 & 53.99 & 45.49 & 54.10 & 55.85 & $100.00 \pm 50.00$ \\
& & Two-shot & 53.99 & 55.21 & 43.51 & 50.81 & 53.77 & $84.62 \pm 38.46$ \\
& & Three-shot & 58.28 & 59.51 & 47.48 & 55.94 & 58.58 & $100.00 \pm 38.46$ \\
\midrule
\multirow{12}{*}{Defects4J}
& \multirow{3}{*}{GPT-4.1 mini}
& One-shot & 14.22 & 14.22 & 12.78 & 14.09 & 14.39 & $84.62 \pm 50.00$ \\
& & Two-shot & 13.05 & 13.29 & 11.69 & 12.81 & 13.29 & $84.62 \pm 56.41$ \\
& & Three-shot & 14.22 & 14.92 & 12.12 & 13.76 & 14.58 & $84.62 \pm 53.85$ \\
\cmidrule(lr){2-9}
& \multirow{3}{*}{Qwen2.5-Coder-32B-Instruct}
& One-shot & 16.32 & 17.25 & 12.89 & 16.15 & 17.29 & $50.00 \pm 71.79$ \\
& & Two-shot & 16.08 & 18.18 & 12.41 & 16.19 & 17.61 & $39.74 \pm 62.82$ \\
& & Three-shot & 15.62 & 17.25 & 12.50 & 16.21 & 17.14 & $39.74 \pm 55.13$ \\
\cmidrule(lr){2-9}
& \multirow{3}{*}{Gemini-2.5-Flash}
& One-shot & 23.78 & 23.78 & 20.23 & 23.65 & 23.78 & $100.00 \pm 51.28$ \\
& & Two-shot & 23.78 & 23.78 & 19.62 & 23.54 & 23.77 & $100.00 \pm 51.28$ \\
& & Three-shot & 22.84 & 23.08 & 20.05 & 23.05 & 23.56 & $84.62 \pm 53.85$ \\
\cmidrule(lr){2-9}
& \multirow{3}{*}{DeepSeek-V3}
& One-shot & 13.75 & 13.75 & 11.26 & 13.50 & 14.21 & $84.62 \pm 60.26$ \\
& & Two-shot & 13.29 & 14.22 & 11.85 & 13.56 & 14.04 & $84.62 \pm 58.97$ \\
& & Three-shot & 13.29 & 13.52 & 11.30 & 12.85 & 13.34 & $84.62 \pm 53.85$ \\
\botrule
\end{tabular}
}
}
\label{tab:few_shot_exact_match}
\end{table*}

\begin{table*}[!htbp]
\centering
\setlength{\tabcolsep}{4pt}
\renewcommand{\arraystretch}{1.08}
\tbl{Partial Match Performance under Few-shot Settings (\%)}
{
\resizebox{\textwidth}{!}{
\begin{tabular}{@{}cccccc@{}}
\toprule
Dataset & Model & Setting & Precision & Recall & F1-score \\
\colrule
\multirow{12}{*}{HumanEval-Java}
& \multirow{3}{*}{GPT-4.1 mini}
& One-shot & $74.81 \pm 0.74$ & $71.93 \pm 0.85$ & $70.17 \pm 0.59$ \\
& & Two-shot & $77.30 \pm 0.33$ & $73.75 \pm 0.45$ & $72.28 \pm 0.31$ \\
& & Three-shot & $75.32 \pm 0.71$ & $74.51 \pm 0.80$ & $71.77 \pm 0.64$ \\
\cmidrule(lr){2-6}
& \multirow{3}{*}{Qwen2.5-Coder-32B-Instruct}
& One-shot & $68.03 \pm 1.02$ & $62.28 \pm 1.01$ & $62.39 \pm 1.12$ \\
& & Two-shot & $70.79 \pm 0.76$ & $62.63 \pm 0.99$ & $63.32 \pm 0.68$ \\
& & Three-shot & $70.99 \pm 0.96$ & $60.75 \pm 1.07$ & $62.59 \pm 0.88$ \\
\cmidrule(lr){2-6}
& \multirow{3}{*}{Gemini-2.5-Flash}
& One-shot & $70.79 \pm 1.81$ & $66.85 \pm 1.76$ & $66.60 \pm 1.67$ \\
& & Two-shot & $73.57 \pm 1.92$ & $70.11 \pm 2.16$ & $69.28 \pm 1.99$ \\
& & Three-shot & $72.95 \pm 1.46$ & $67.71 \pm 1.80$ & $67.63 \pm 1.59$ \\
\cmidrule(lr){2-6}
& \multirow{3}{*}{DeepSeek-V3}
& One-shot & $69.74 \pm 1.00$ & $66.89 \pm 2.53$ & $65.44 \pm 1.08$ \\
& & Two-shot & $69.82 \pm 1.31$ & $71.13 \pm 2.33$ & $66.92 \pm 1.40$ \\
& & Three-shot & $72.73 \pm 1.06$ & $70.58 \pm 2.24$ & $68.33 \pm 1.43$ \\
\midrule
\multirow{12}{*}{Defects4J}
& \multirow{3}{*}{GPT-4.1 mini}
& One-shot & $29.84 \pm 0.50$ & $26.04 \pm 0.74$ & $23.32 \pm 0.42$ \\
& & Two-shot & $30.50 \pm 0.48$ & $32.69 \pm 0.69$ & $25.65 \pm 0.35$ \\
& & Three-shot & $30.13 \pm 0.73$ & $36.39 \pm 0.72$ & $25.77 \pm 0.60$ \\
\cmidrule(lr){2-6}
& \multirow{3}{*}{Qwen2.5-Coder-32B-Instruct}
& One-shot & $25.62 \pm 0.91$ & $20.76 \pm 0.78$ & $19.92 \pm 0.64$ \\
& & Two-shot & $24.95 \pm 0.96$ & $24.37 \pm 0.63$ & $19.84 \pm 0.64$ \\
& & Three-shot & $25.80 \pm 0.63$ & $28.05 \pm 1.33$ & $21.13 \pm 0.58$ \\
\cmidrule(lr){2-6}
& \multirow{3}{*}{Gemini-2.5-Flash}
& One-shot & $43.25 \pm 0.46$ & $32.74 \pm 0.52$ & $32.90 \pm 0.52$ \\
& & Two-shot & $43.97 \pm 1.81$ & $34.17 \pm 1.01$ & $33.60 \pm 1.29$ \\
& & Three-shot & $43.54 \pm 0.63$ & $35.33 \pm 0.73$ & $34.40 \pm 0.66$ \\
\cmidrule(lr){2-6}
& \multirow{3}{*}{DeepSeek-V3}
& One-shot & $25.61 \pm 0.77$ & $26.57 \pm 0.89$ & $20.44 \pm 0.66$ \\
& & Two-shot & $26.74 \pm 0.66$ & $28.20 \pm 1.05$ & $21.35 \pm 0.44$ \\
& & Three-shot & $24.92 \pm 0.61$ & $30.34 \pm 0.99$ & $21.29 \pm 0.52$ \\
\botrule
\end{tabular}
}
}
\label{tab:few_shot_partial_match}
\end{table*}

\noindent
\fbox{%
\begin{minipage}{\dimexpr\textwidth-2\fboxsep-2\fboxrule}
\textbf{Finding 1}: Few-shot prompting more consistently improves Recall on Defects4J than on HumanEval-Java, suggesting that in-context examples may help models cover more ground-truth buggy lines in complex project settings.

\medskip
\textbf{Finding 2}: Increasing the number of in-context examples does not lead to monotonic improvements in exact-match localization. One-shot, Two-shot, and Three-shot settings achieve the best EM-based performance for different models and datasets, indicating that the marginal benefit of Few-shot prompting is model- and task-dependent.

\medskip
\textbf{Finding 3}: Few-shot prompting can improve output consistency for some models. Compared with Zero-shot prompting, Few-shot settings generally lead to higher or more concentrated CR values for several models, suggesting that in-context examples may constrain the model's localization pattern.
\end{minipage}
}

\subsection{RQ3: Effect of Chain-of-Thought Prompting}
\textbf{[Experiment Goal]}: This experiment aims to evaluate the effect of Chain-of-Thought (CoT) prompting on the statement-level fault localization performance of LLMs. By comparing the localization performance of four LLMs under the CoT and Zero-shot settings, we analyze the effect of explicit step-by-step reasoning guidance on model localization results.

\textbf{[Experiment Design]}: We extend the Zero-shot prompt with CoT-style reasoning guidance and evaluate the four LLMs on the HumanEval-Java and Defects4J datasets. For each bug case, we perform 13 independent API calls under the same prompt and API parameter settings, and compute model performance according to the evaluation metrics defined in Section~3.5. For HumanEval-Java, the CoT prompts are designed with step-by-step reasoning guidance based only on source code. For Defects4J, the CoT prompts further incorporate bug report context, guiding the model to identify potential buggy lines based on error symptoms and code logic. The complete CoT prompt templates are provided in Appendix~B.

\textbf{[Experimental Results]}: As shown in Table~8, CoT prompting does not yield consistent improvement on HumanEval-Java. Compared with the Zero-shot setting, the Pass@10 values of GPT-4.1 mini, Qwen2.5-Coder-32B-Instruct, and DeepSeek-V3 decrease to 35.25\%, 47.37\%, and 49.85\%, respectively. Gemini-2.5-Flash still maintains a relatively high Pass@10 of 67.27\%, but this is also slightly lower than its Zero-shot result of 68.73\%. In terms of CR, DeepSeek-V3 achieves a \(CR_{\mathrm{median}}\) of 100.00\%, but its localization performance still decreases, indicating that higher output consistency does not necessarily imply better localization effectiveness.

On Defects4J, the effect of CoT is more model-dependent. Compared with the Zero-shot setting with bug report context, the Pass@10 values of Qwen2.5-Coder-32B-Instruct and DeepSeek-V3 increase to 16.63\% and 19.07\%, respectively. In contrast, the Pass@10 values of GPT-4.1 mini and Gemini-2.5-Flash decrease to 15.02\% and 18.88\%, respectively. These results suggest that CoT can benefit some models in fault localization for real-world projects, but it does not guarantee consistent improvement across all models.

As shown in Table~9, the effect of CoT on PM-based metrics is also limited. On HumanEval-Java, Gemini-2.5-Flash and DeepSeek-V3 achieve relatively high F1-scores of 67.81\% and 66.47\%, respectively, whereas GPT-4.1 mini decreases markedly to 31.88\%. On Defects4J, the F1-scores of Qwen2.5-Coder-32B-Instruct and DeepSeek-V3 increase to 18.84\% and 24.04\%, respectively, while those of GPT-4.1 mini and Gemini-2.5-Flash are slightly lower than their Zero-shot results with bug report context. Overall, CoT may help certain models organize the reasoning process and identify relevant buggy lines, but it is not a universally effective enhancement strategy for fault localization.

\begin{table*}[!htbp]
\centering
\setlength{\tabcolsep}{4pt}
\renewcommand{\arraystretch}{1.12}
\tbl{Exact Match Performance under the Chain-of-Thought Setting (\%)}
{
\resizebox{\textwidth}{!}{
\begin{tabular}{@{}cccccccc@{}}
\toprule
Dataset & Model & Top@5 & Top@10 & Pass@1 & Pass@5 & Pass@10 & CR$_{median}$ $\pm$ IQR \\
\colrule
\multirow{4}{*}{HumanEval-Java}
& GPT-4.1 mini & 33.74 & 34.97 & 22.37 & 32.26 & 35.25 & $84.62 \pm 46.15$ \\
& Qwen2.5-Coder-32B-Instruct & 46.63 & 47.24 & 36.24 & 45.54 & 47.37 & $84.62 \pm 51.28$ \\
& Gemini-2.5-Flash & 63.19 & 66.26 & 50.64 & 63.19 & 67.27 & $84.62 \pm 51.28$ \\
& DeepSeek-V3 & 47.85 & 50.31 & 44.31 & 48.69 & 49.85 & $100.00 \pm 51.28$ \\
\midrule
\multirow{4}{*}{Defects4J}
& GPT-4.1 mini & 14.69 & 15.15 & 13.14 & 14.56 & 15.02 & $84.62 \pm 51.28$ \\
& Qwen2.5-Coder-32B-Instruct & 16.08 & 17.02 & 10.72 & 14.81 & 16.63 & $53.85 \pm 57.34$ \\
& Gemini-2.5-Flash & 18.88 & 18.88 & 16.68 & 18.80 & 18.88 & $71.79 \pm 53.85$ \\
& DeepSeek-V3 & 19.11 & 19.58 & 14.60 & 17.92 & 19.07 & $70.51 \pm 60.26$ \\
\botrule
\end{tabular}
}
}
\label{tab:cot_exact_match}
\end{table*}

\begin{table*}[!htbp]
\centering
\setlength{\tabcolsep}{5pt}
\renewcommand{\arraystretch}{1.12}
\tbl{Partial Match Performance under the Chain-of-Thought Setting (\%)}
{
\resizebox{\textwidth}{!}{
\begin{tabular}{@{}ccccc@{}}
\toprule
Dataset & Model & Precision & Recall & F1-score \\
\colrule
\multirow{4}{*}{HumanEval-Java}
& GPT-4.1 mini & $35.45 \pm 2.82$ & $30.99 \pm 2.64$ & $31.88 \pm 2.65$ \\
& Qwen2.5-Coder-32B-Instruct & $62.09 \pm 1.60$ & $58.47 \pm 1.93$ & $57.32 \pm 1.55$ \\
& Gemini-2.5-Flash & $72.45 \pm 1.51$ & $67.83 \pm 1.83$ & $67.81 \pm 1.47$ \\
& DeepSeek-V3 & $67.95 \pm 1.16$ & $74.35 \pm 0.89$ & $66.47 \pm 0.93$ \\
\midrule
\multirow{4}{*}{Defects4J}
& GPT-4.1 mini & $30.03 \pm 0.67$ & $29.78 \pm 1.27$ & $23.45 \pm 0.71$ \\
& Qwen2.5-Coder-32B-Instruct & $24.44 \pm 0.81$ & $20.23 \pm 1.09$ & $18.84 \pm 0.80$ \\
& Gemini-2.5-Flash & $38.64 \pm 0.87$ & $32.17 \pm 1.50$ & $30.95 \pm 0.90$ \\
& DeepSeek-V3 & $31.42 \pm 0.94$ & $25.48 \pm 0.87$ & $24.04 \pm 0.70$ \\
\botrule
\end{tabular}
}
}
\label{tab:cot_partial_match}
\end{table*}

\medskip
\noindent
\fbox{%
\begin{minipage}{\dimexpr\textwidth-2\fboxsep-2\fboxrule}
\textbf{Finding 1}: CoT prompting is not universally beneficial for statement-level fault localization. For relatively simple or controlled faults, it may introduce unnecessary reasoning overhead; for more complex project-level faults, it can help some models leverage contextual clues, but the effect remains model-dependent.

\end{minipage}
}

\subsection{RQ4: End-to-end Response Time and API Cost}

This study evaluates the practical efficiency and cost of LLM-based fault localization from the perspectives of time efficiency and token-based API cost, providing empirical evidence for model selection in industrial scenarios.
\subsubsection{End-to-end Response Time Analysis}
\textbf{[Experiment Goal]}: This experiment aims to evaluate the time efficiency of different LLMs in statement-level fault localization and analyze how model type and prompt strategy affect end-to-end response time. The goal is to provide efficiency-oriented evidence for model selection in practical applications.

\textbf{[Experiment Design]}: We measure the end-to-end response time of the four LLMs under different prompt settings, including Zero-shot, Few-shot, and CoT. For Defects4J, we further distinguish between the Zero-shot settings without and with bug report context. For each bug case, we perform 13 independent API calls under the same prompt and API parameter settings, and record the elapsed time from sending the API request to receiving the complete response. We then group response times by localization success and failure, visualize the distributions using boxplots, and apply the Scott-Knott ESD test to analyze differences across models and settings \cite{pornprasitDeepLineDPDeepLearning2023,qiuBAFLineDPCodeBilinear2024}.

\textbf{[Experimental Results]}: As shown in Figure~1, on HumanEval-Java, the four models exhibit clear differences in end-to-end response time. GPT-4.1 mini and Qwen2.5-Coder-32B-Instruct achieve the shortest response times, with average values of approximately 1--2 seconds, and their response times increase only slightly as the number of Few-shot examples increases. DeepSeek-V3 shows relatively stable average response times of approximately 7--11 seconds. Gemini-2.5-Flash has the longest average response time, with some responses exceeding 30 seconds under the CoT setting, suggesting that CoT-style reasoning introduces additional response latency for this model.

As shown in Figure~2, on Defects4J, the overall end-to-end response time of the four models is higher than that on HumanEval-Java because of the larger code inputs. GPT-4.1 mini still maintains high time efficiency, with an average response time of approximately 1--3 seconds. Qwen2.5-Coder-32B-Instruct remains relatively stable overall, with an average response time of approximately 2--5 seconds; however, under the Zero-shot setting without bug report context, the average response time for failed localization cases increases to 14.04 seconds. DeepSeek-V3 has an average response time of approximately 3--9 seconds, and under the Zero-shot setting without bug report context, the average response time for failed localization cases reaches 22.96 seconds. Gemini-2.5-Flash shows the longest overall response time, with successful localization cases taking about 9 seconds and failed localization cases usually exceeding 20 seconds. Overall, bug report context is associated not only with improved localization performance, but also with shorter response times for completing fault localization in most models.

\begin{figure*}[t]
\centering
\begin{minipage}[b]{0.45\textwidth}
    \centering
    \includegraphics[width=\linewidth]{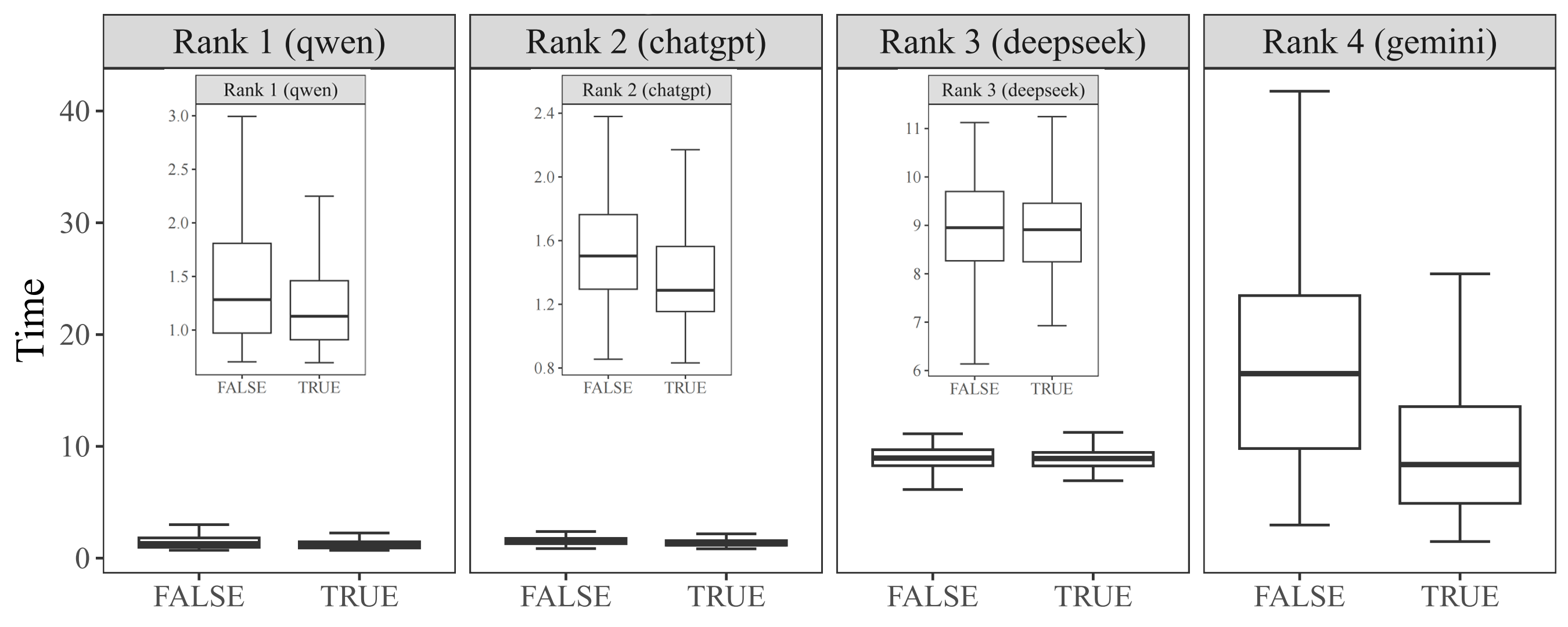}
    \vspace{0.3em}
    \textbf{(a)} Zero-shot
\end{minipage}
\hfill
\begin{minipage}[b]{0.45\textwidth}
    \centering
    \includegraphics[width=\linewidth]{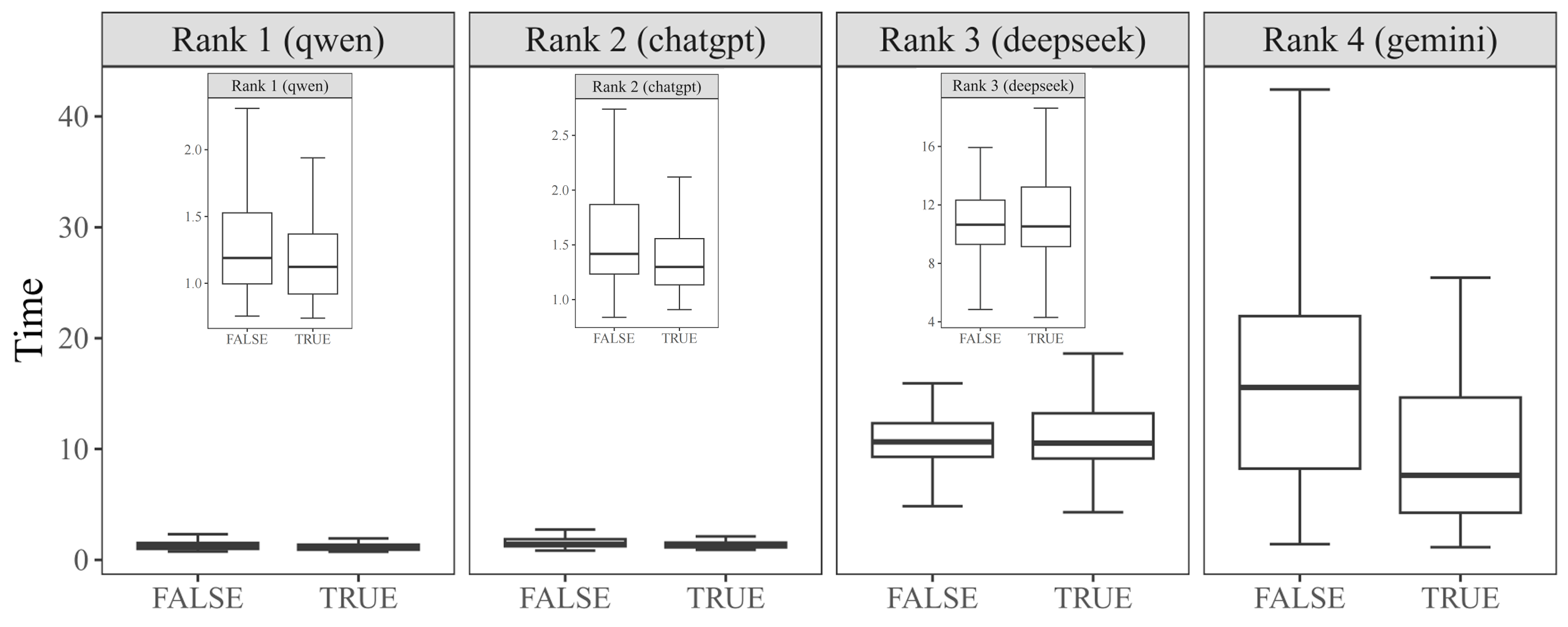}
    \vspace{0.3em}
    \textbf{(b)} One-shot
\end{minipage}

\vspace{1.5em}

\begin{minipage}[b]{0.45\textwidth}
    \centering
    \includegraphics[width=\linewidth]{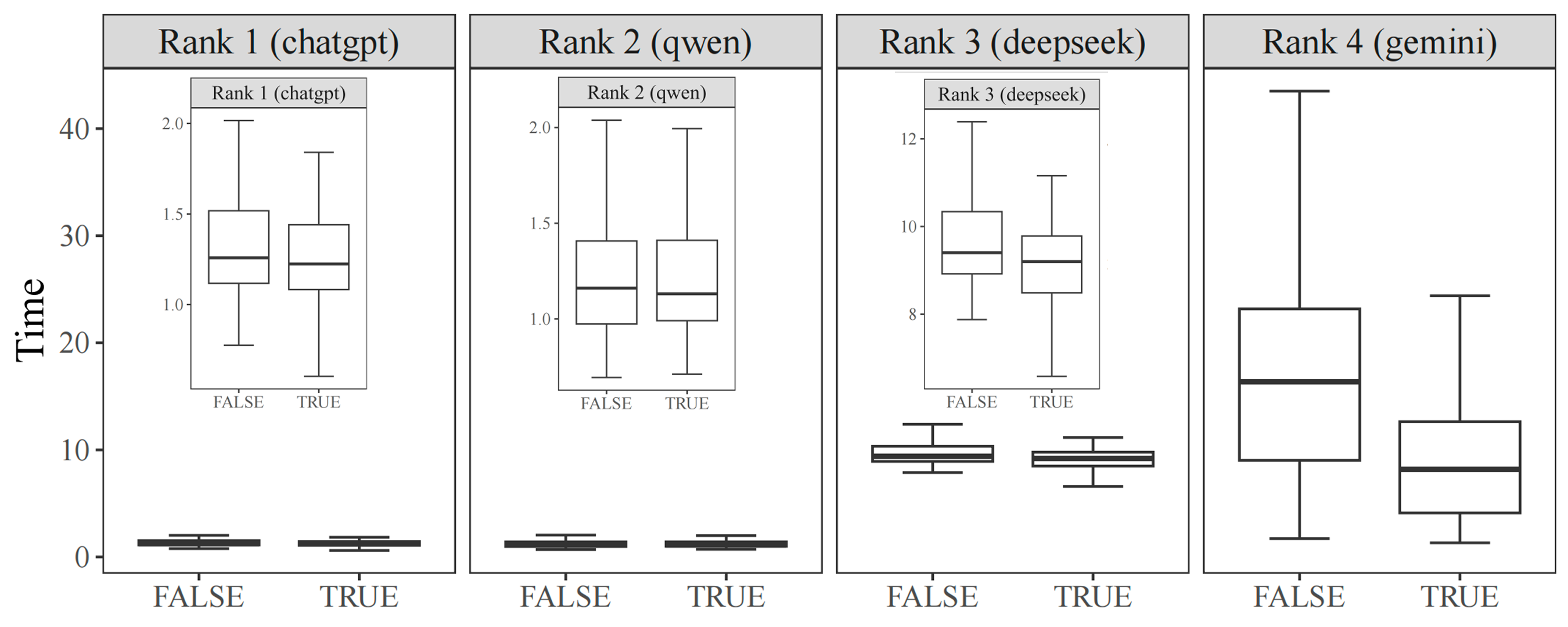}
    \vspace{0.3em}
    \textbf{(c)} Two-shot
\end{minipage}
\hfill
\begin{minipage}[b]{0.45\textwidth}
    \centering
    \includegraphics[width=\linewidth]{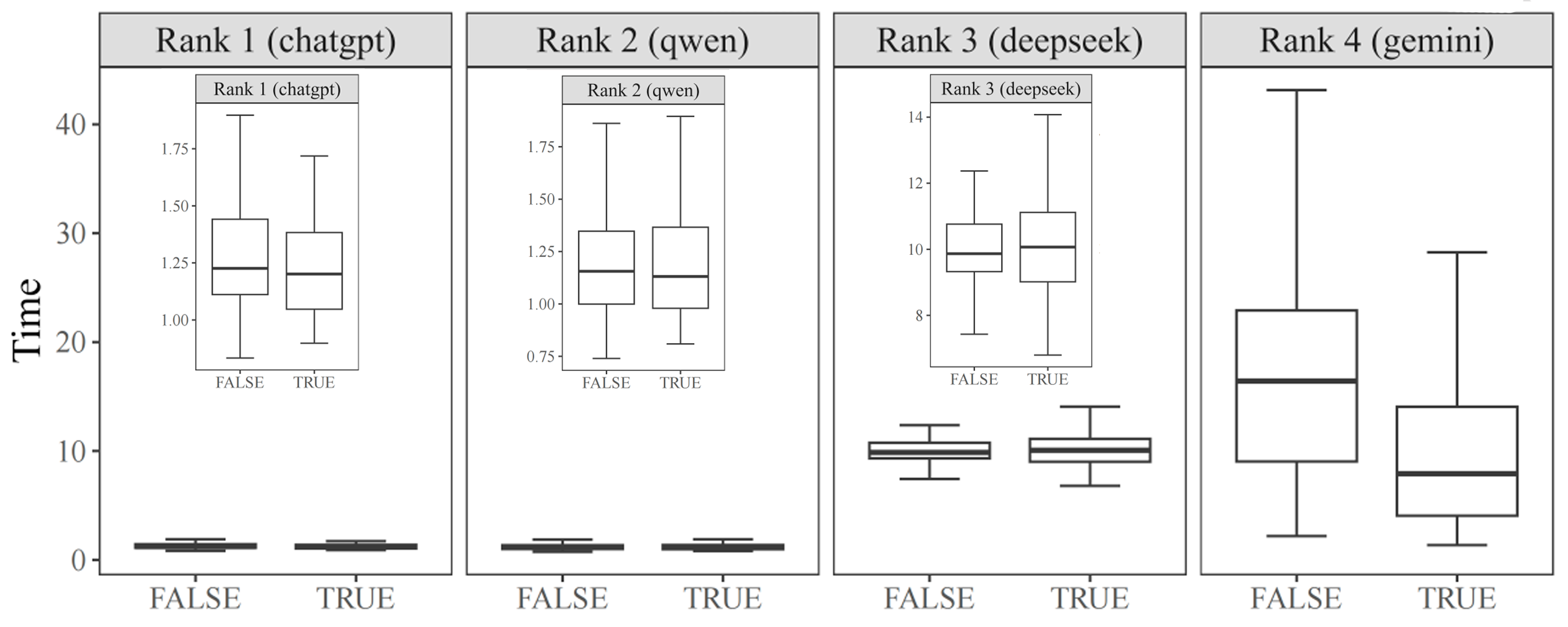}
    \vspace{0.3em}
    \textbf{(d)} Three-shot
\end{minipage}

\vspace{1.5em}

\begin{minipage}[b]{0.45\textwidth}
    \centering
    \includegraphics[width=\linewidth]{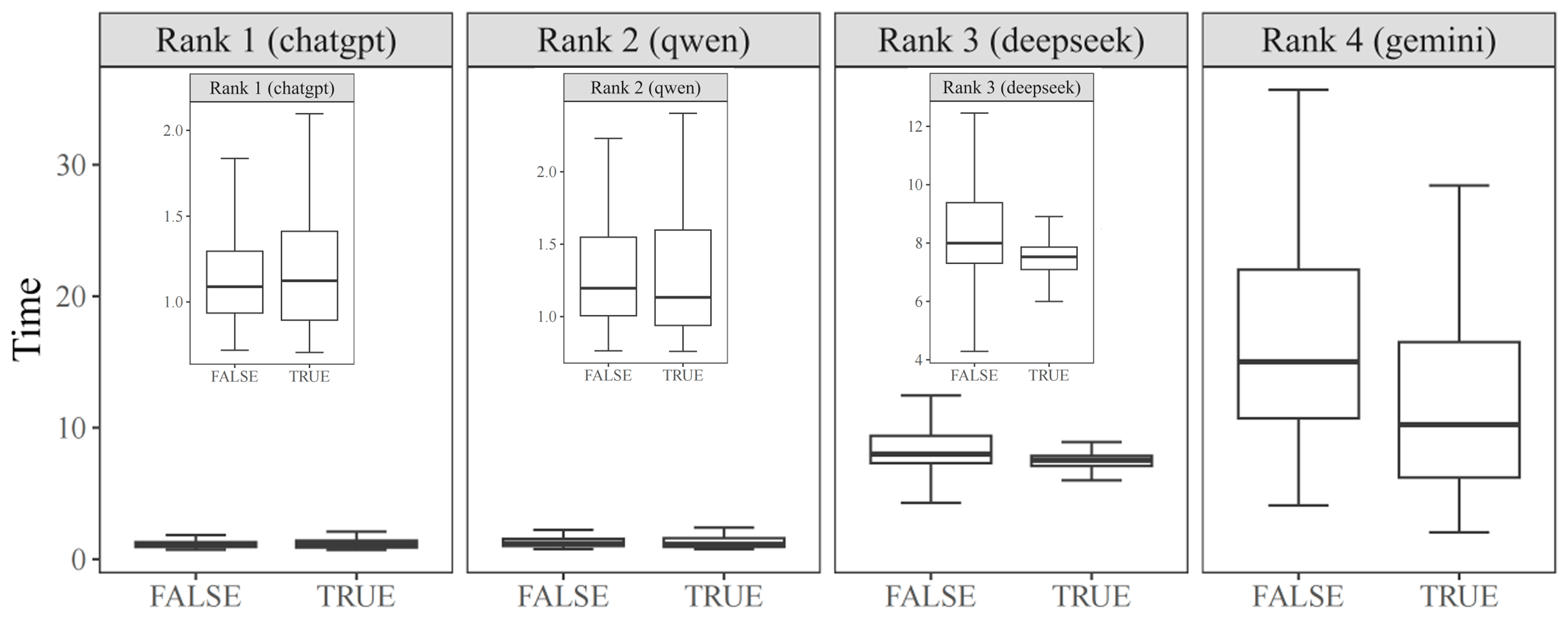}
    \vspace{0.3em}
    \textbf{(e)} CoT
\end{minipage}

\caption{End-to-end response time on HumanEval-Java dataset.}
\label{fig:time_humaneval}
\end{figure*}

\begin{figure*}[t]
\centering

\begin{minipage}[b]{0.45\textwidth}
    \centering
    \includegraphics[width=\linewidth]{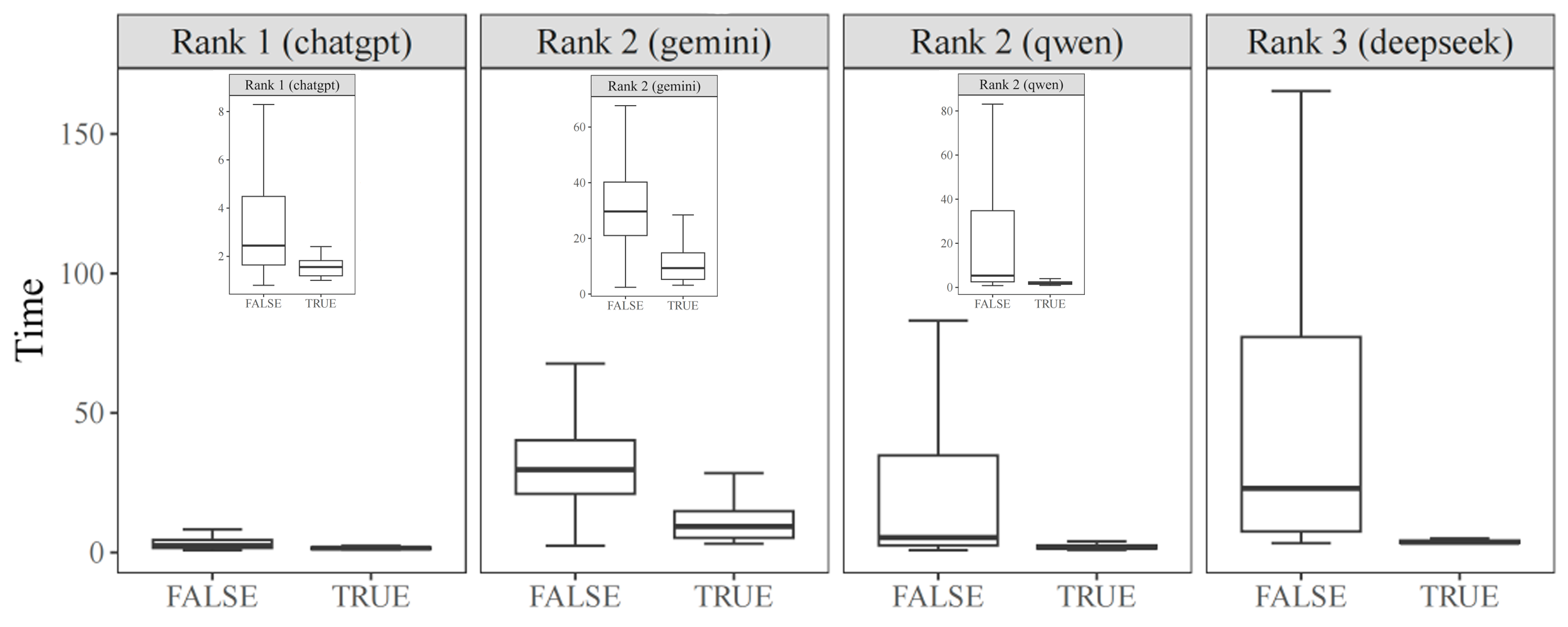}
    \vspace{0.3em}
    \textbf{(a)} Zero-shot w/o bug report
\end{minipage}
\hfill
\begin{minipage}[b]{0.45\textwidth}
    \centering
    \includegraphics[width=\linewidth]{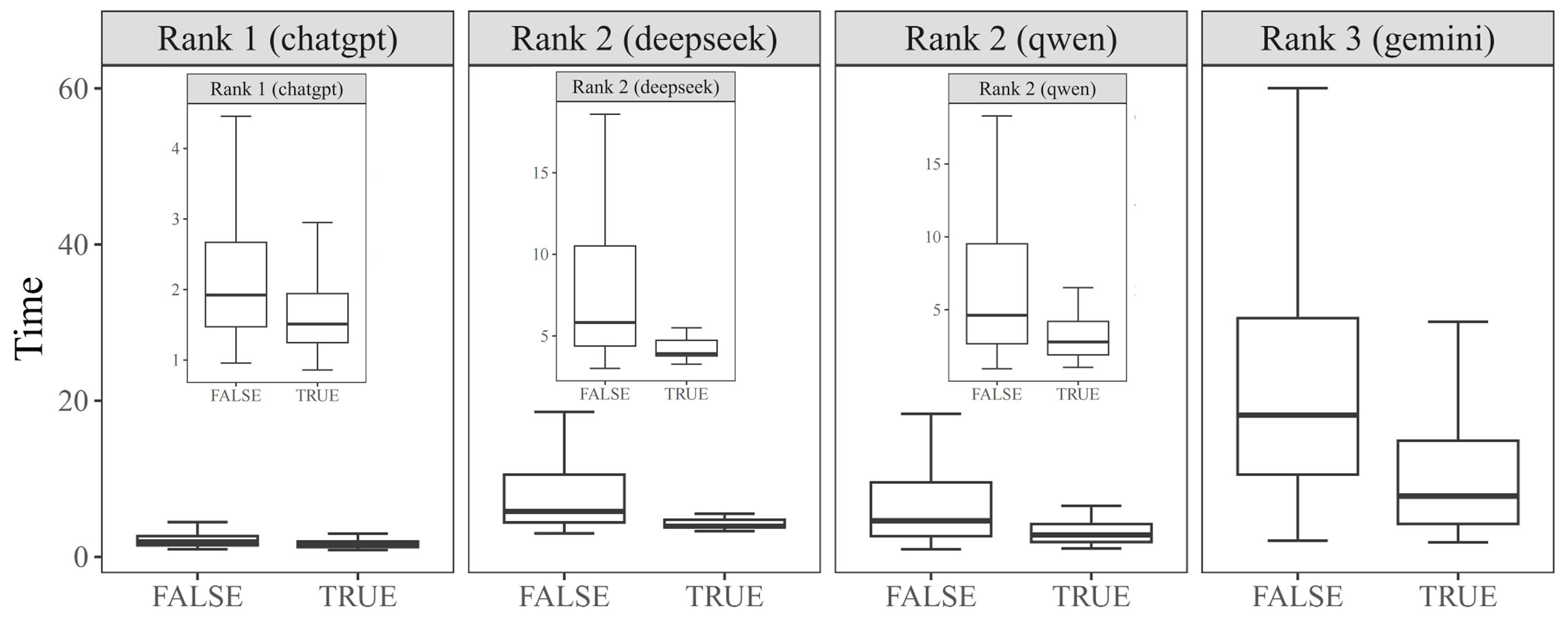}
    \vspace{0.3em}
    \textbf{(b)} Zero-shot w/ bug report
\end{minipage}

\vspace{1.5em}

\begin{minipage}[b]{0.45\textwidth}
    \centering
    \includegraphics[width=\linewidth]{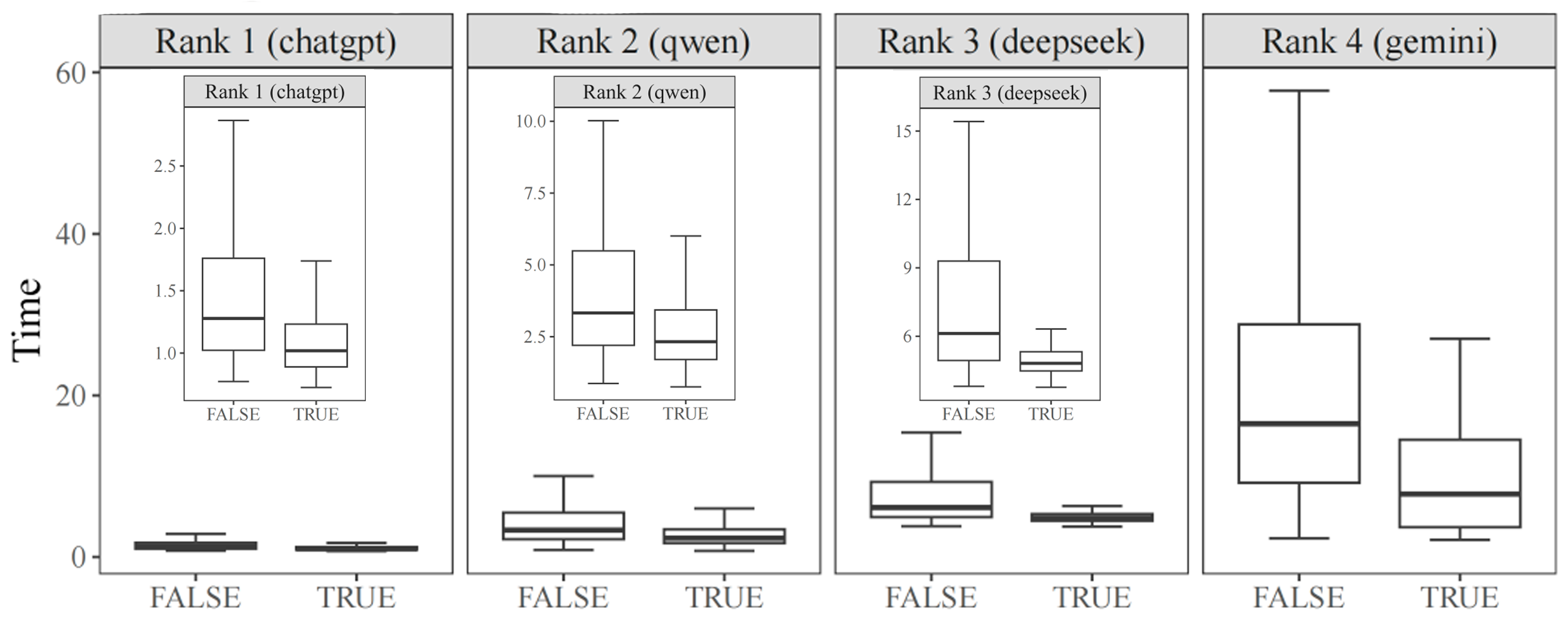}
    \vspace{0.3em}
    \textbf{(c)} One-shot
\end{minipage}
\hfill
\begin{minipage}[b]{0.45\textwidth}
    \centering
    \includegraphics[width=\linewidth]{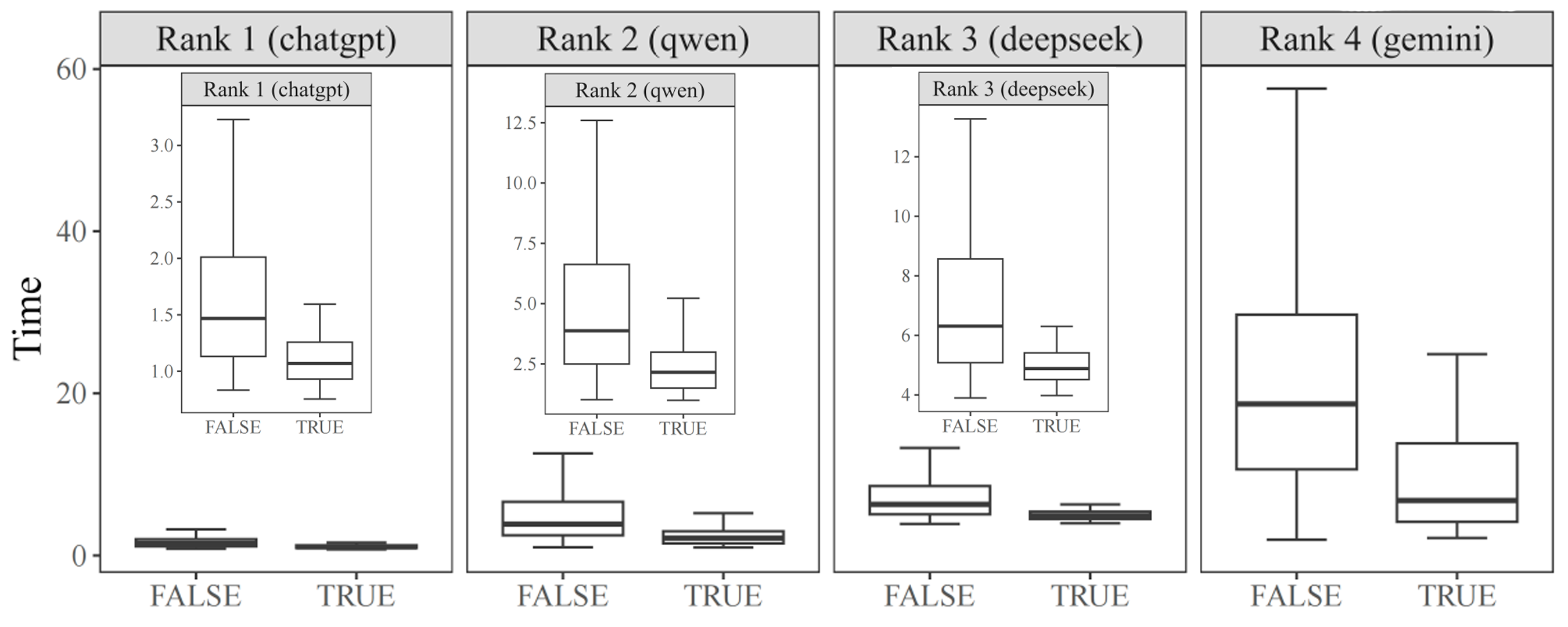}
    \vspace{0.3em}
    \textbf{(d)} Two-shot
\end{minipage}

\vspace{1.5em}

\begin{minipage}[b]{0.45\textwidth}
    \centering
    \includegraphics[width=\linewidth]{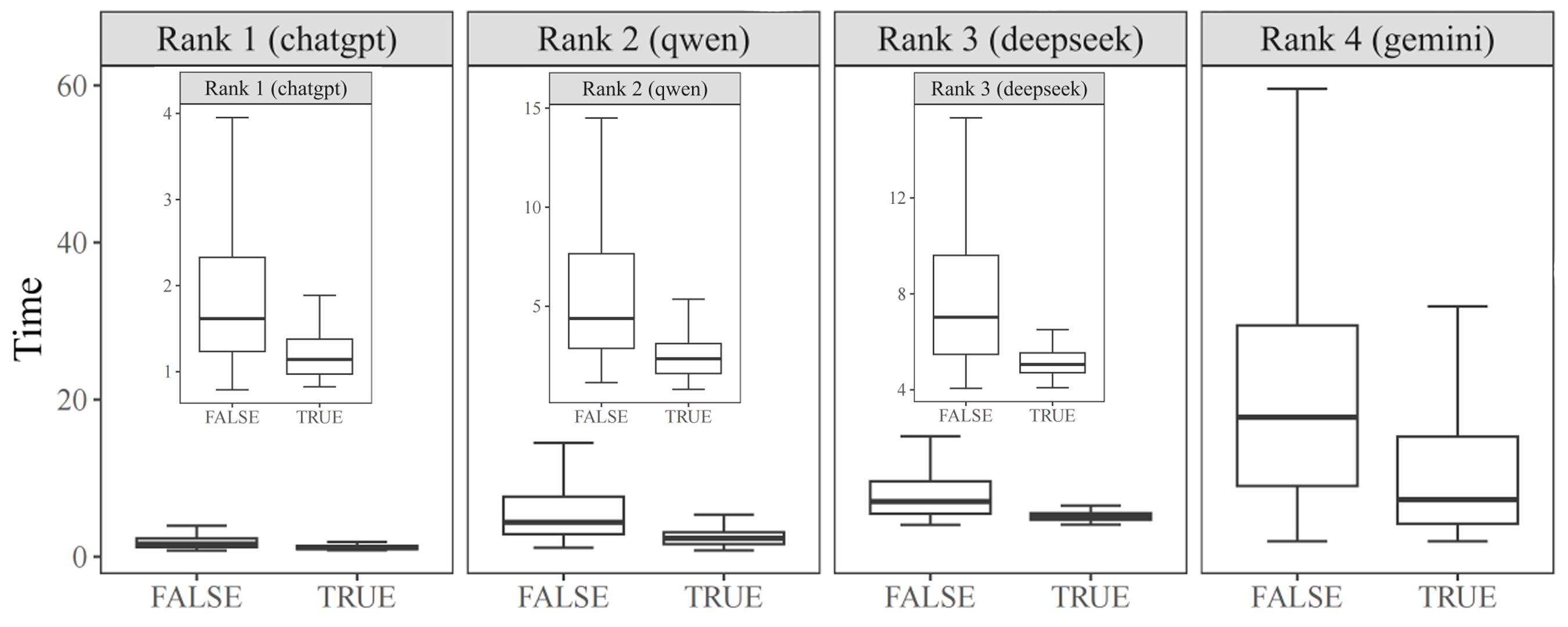}
    \vspace{0.3em}
    \textbf{(e)} Three-shot
\end{minipage}
\hfill
\begin{minipage}[b]{0.45\textwidth}
    \centering
    \includegraphics[width=\linewidth]{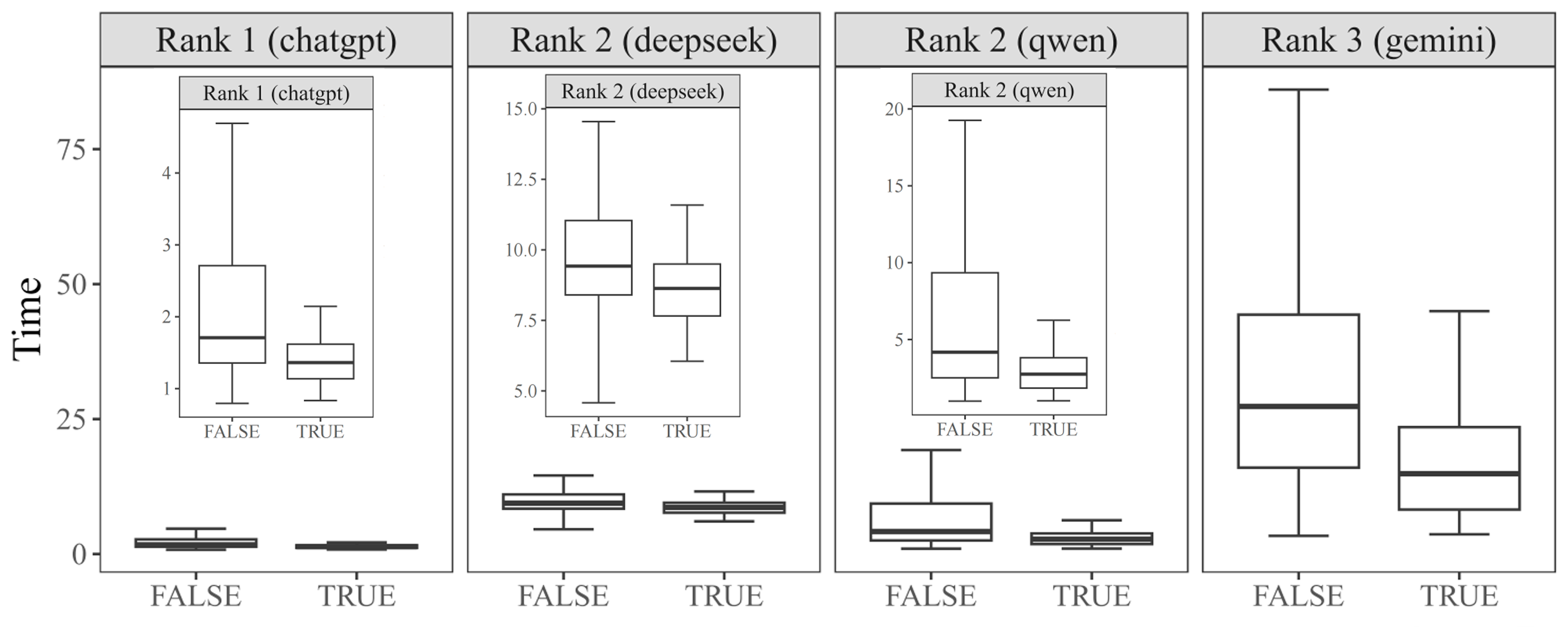}
    \vspace{0.3em}
    \textbf{(f)} CoT
\end{minipage}

\caption{End-to-end response time on Defects4J dataset.}
\label{fig:time_defects4j}
\end{figure*}

\noindent
\fbox{%
\begin{minipage}{\dimexpr\textwidth-2\fboxsep-2\fboxrule}
\textbf{Finding 1}: End-to-end response time is associated with localization difficulty. Failed localization cases generally incur higher latency, particularly on Defects4J, indicating that complex project-level faults impose greater reasoning and generation overhead on LLMs.

\medskip
\textbf{Finding 2}: Context length alone does not determine response efficiency; context relevance matters more. On Defects4J, adding bug report context increases input length but is associated with lower observed end-to-end response time for most models, suggesting that task-relevant diagnostic information can help constrain the localization search space.
\end{minipage}
}

\subsubsection{API Cost Analysis}

\textbf{[Experiment Goal]}: This experiment aims to evaluate the token-based API cost of different LLMs in statement-level fault localization and analyze their monetary overhead in practical applications.

\textbf{[Experiment Design]}: We use the Zero-shot prompting setting without bug report context. For each model, a full single-call evaluation denotes querying each bug case once in a dataset. Since each bug case is queried 13 times in the repeated-call experiments, we compute the input tokens, output tokens, and API cost for each full single-call evaluation and report the average over the 13 repeated evaluations. The API cost is calculated according to the official token pricing provided by each model provider, and all total costs are converted into USD.

\textbf{[Experimental Results]}: As shown in Table~10, on HumanEval-Java, the numbers of output tokens generated by different models are generally comparable, and therefore the differences in total cost are relatively small. Qwen2.5-Coder-32B-Instruct incurs the lowest cost, at \$0.0200, whereas GPT-4.1 mini incurs the highest cost, at \$0.0322.

On Defects4J, because of the larger code inputs, the total costs of all models are much higher than those on HumanEval-Java. In addition, the numbers of output tokens vary substantially across models: Gemini-2.5-Flash generates the fewest output tokens, with 26,199 tokens, whereas DeepSeek-V3 generates the most, with 434,603 tokens. The corresponding cost results show that Qwen2.5-Coder-32B-Instruct incurs the lowest cost, at \$1.2003, while GPT-4.1 mini incurs the highest cost, at \$1.6553. Overall, API cost is mainly affected by dataset size, input and output token counts, and the pricing strategies of model providers.

\begin{table*}[!htbp]
\centering
\setlength{\tabcolsep}{4pt}
\renewcommand{\arraystretch}{1.12}
\tbl{API Cost Comparison of Different Models on Two Datasets under the Zero-shot Setting}
{
\resizebox{\textwidth}{!}{
\begin{tabular}{@{}ccccccc@{}}
\toprule
Dataset & Model & Input Tokens & Output Tokens & Input Price & Output Price & Total Cost (\$) \\
\colrule
\multirow{4}{*}{HumanEval-Java}
& GPT-4.1 mini & 55546 & 6250 & \$0.4/1M Tokens & \$1.6/1M Tokens & 0.0322 \\
& Qwen2.5-Coder-32B-Instruct & 55880 & 5406 & \textyen0.002/1K Tokens & \textyen0.006/1K Tokens & 0.0200 \\
& Gemini-2.5-Flash & 60926 & 5060 & \$0.30/1M Tokens & \$2.5/1M Tokens & 0.0310 \\
& DeepSeek-V3 & 55074 & 5415 & \textyen2/1M Tokens & \textyen8/1M Tokens & 0.0213 \\
\midrule
\multirow{4}{*}{Defects4J}
& GPT-4.1 mini & 3552461 & 146462 & \$0.4/1M Tokens & \$1.6/1M Tokens & 1.6553 \\
& Qwen2.5-Coder-32B-Instruct & 3531194 & 263221 & \textyen0.002/1K Tokens & \textyen0.006/1K Tokens & 1.2003 \\
& Gemini-2.5-Flash & 4037135 & 26199 & \$0.30/1M Tokens & \$2.5/1M Tokens & 1.2770 \\
& DeepSeek-V3 & 3532065 & 434603 & \textyen2/1M Tokens & \textyen8/1M Tokens & 1.4640 \\
\botrule
\end{tabular}
}
}
\par\smallskip
\noindent\parbox{\textwidth}{\raggedright\footnotesize Note: Token counts and total costs are averaged over 13 repeated full single-call evaluations under the Zero-shot source-code-only setting. Prices are shown in the original billing currencies, while Total Cost is reported in USD using USD 1 = CNY 7.20. \par}
\label{tab:api_cost}
\end{table*}

\noindent
\fbox{%
\begin{minipage}{\dimexpr\textwidth-2\fboxsep-2\fboxrule}
\textbf{Finding 1}: API cost and fault localization performance do not exhibit a simple positive relationship. In the Zero-shot source-code-only setting, Gemini-2.5-Flash achieves the highest observed EM-based performance and the highest PM-based F1-score among the evaluated models on Defects4J, while generating the fewest output tokens on both datasets. This indicates a favorable effectiveness-efficiency trade-off.

\medskip
\textbf{Finding 2}: Output verbosity does not directly characterize localization capability. On Defects4J, models generating more output tokens do not show better EM or PM performance. This suggests that concise, task-aligned outputs are more cost-effective for statement-level fault localization.

\end{minipage}
}

%% file: 5.threats_to_validity.tex
\section{Threats to Validity}\label{sec5}
\textbf{Threats to Internal Validity}: The internal validity of this study is primarily influenced by two factors: First, the implementation of prompt engineering may introduce systematic errors. While standardized templates for the Zero-shot, Few-shot, and CoT strategies were developed, the choice of examples in the prompts, the specific wording of the prompt templates, and the guidance of Chain-of-Thought reasoning may still affect model outputs. To minimize these biases, all experiments followed a unified prompt structure, and the Few-shot examples were carefully selected to ensure representativeness and neutrality. Second, external factors such as fluctuations in API performance, including network delays and server load, could affect the results. To mitigate this, we conducted 13 independent API calls for each bug case on the same device and under the same API parameter settings, and computed metrics such as Top@k, Pass@k, and CR to characterize the models' localization performance and output consistency across repeated calls.

\textbf{Threats to Construct Validity}: The construct validity of this study is primarily threatened by the following four factors: First, although metrics such as Top@k, Pass@k, CR, Precision, Recall, F1-score, and API cost evaluate model performance from the perspectives of localization accuracy, output consistency, and application cost, they still cannot cover all capabilities involved in fault localization. Second, model outputs need to be mapped to source-code lines using unified rules. Formatting deviations, incomplete statement fragments, or inconsistent line numbers in the model output may cause some partially reasonable localization results to be misjudged. Third, the main experiments in this study primarily evaluate the statement-level fault localization capability of LLMs under text-based input settings, where the inputs mainly consist of source-code text and bug report context. However, real-world debugging often relies on multiple sources of information, such as program behavior, test results, and structured program representations. Therefore, the results of this study cannot fully reflect the performance upper bound of LLMs when richer debugging information is available. Fourth, since the release time and knowledge cutoff of some models are later than the release dates of the datasets used in this study, benchmark exposure cannot be entirely ruled out, which may affect the independence and reliability of the evaluation results.

\textbf{Threats to External Validity}: The external validity of this study is primarily threatened by two aspects: First, the experiments were conducted only in the Java programming language and on two datasets, HumanEval-Java and Defects4J, meaning that the applicability of the conclusions to other programming languages, project sizes, and complex fault scenarios remains to be further verified. Second, although this study selected four representative LLMs and two non-LLM baselines, it still cannot cover all models and fault localization methods. Moreover, model iterations, API pricing strategies, server load, and changes in network conditions may also affect the long-term applicability of the time and cost results.

%% file: 6.conclusions.tex
\section{Conclusions}\label{sec6}
This study presents a systematic empirical evaluation of LLM-based statement-level fault localization. Using HumanEval-Java and Defects4J as benchmarks, we evaluate four representative LLMs, GPT-4.1 mini, Gemini-2.5-Flash, Qwen2.5-Coder-32B-Instruct, and DeepSeek-V3, and compare them with two non-LLM baselines, PMD and LineDef. The results show that LLMs exhibit different strengths and limitations across datasets, input contexts, prompt strategies, and buggy-line counts. Bug report context improves observed localization performance on Defects4J, especially for simpler bugs, whereas complex multi-line faults remain challenging. Few-shot prompting improves performance in some cases, but more examples do not consistently lead to higher exact-match success. Chain-of-Thought prompting shows model-dependent effects and does not provide universal improvements. Moreover, Exact Match, Partial Match, and Consistency Rate capture different aspects of model behavior, suggesting that LLM-based fault localization should be evaluated from multiple complementary perspectives. The analysis of end-to-end response time and token-based API cost further highlights the need to consider both effectiveness and practical cost when selecting models.

Future work will extend the evaluation framework by incorporating richer debugging information, such as test execution results, program behavior, and structured program representations. We will also explore more effective reasoning and context-selection mechanisms for complex multi-line and cross-module faults. In addition, we plan to expand the evaluation to more programming languages, project types, and fault categories, and continuously track the performance of emerging LLMs and non-LLM fault localization methods.

%% file: 7.Appendix.tex
\section{Output Variability under Temperature = 0}
\label{sec:output_variability}

\begin{table}[H]
\centering
\setlength{\tabcolsep}{8pt}
\renewcommand{\arraystretch}{1.12}
\tbl{Output variability under temperature = 0 in the Zero-shot source-code-only setting}
{
\begin{tabular}{@{}cccc@{}}
\toprule
Dataset & Model & Cases with non-identical outputs & Average unique outputs per case \\
\colrule
\multirow{4}{*}{HumanEval-Java}
& GPT-4.1 mini & 42.33\% & 1.72 \\
& Qwen2.5-Coder-32B-Instruct & 56.44\% & 2.67 \\
& Gemini-2.5-Flash & 65.03\% & 2.39 \\
& DeepSeek-V3 & 47.24\% & 1.94 \\
\midrule
\multirow{4}{*}{Defects4J}
& GPT-4.1 mini & 70.86\% & 3.82 \\
& Qwen2.5-Coder-32B-Instruct & 86.95\% & 4.74 \\
& Gemini-2.5-Flash & 82.05\% & 1.82 \\
& DeepSeek-V3 & 78.55\% & 5.48 \\
\botrule
\end{tabular}
}
\par\smallskip
\noindent\makebox[\textwidth][l]{\parbox{\textwidth}{\raggedright\footnotesize Note: Results are reported under the Zero-shot source-code-only setting. “Non-identical outputs” denotes cases with two or more distinct localized buggy-line sets across 13 runs; “Average unique outputs” denotes the mean number of distinct localized buggy-line sets per case.\par}}
\label{tab:output_variability}
\end{table}

\section{Prompt Templates}\label{sec7}
\begin{figure}[htbp]
\begin{topbox}
    Analyze the provided Java code and identify the exact line numbers and statements that contain bugs. Return the exact buggy code lines and bug statements as they are without any explanations or code repair advice. Your response must strictly follow the format: 'Line X:bug statement'. Do not provide any additional information or suggestions.

    Java code:

    [Java Source Code for Fault Localization]
\end{topbox}

\begin{bottombox}
    bug report:

    [Bug Report for Java Source Code]
\end{bottombox}
\raggedright\caption{Prompt template for Zero-shot fault localization.}
\label{fig:prompt-Zero-Shot}
\end{figure}

\begin{figure}[htbp]
\centering
\begin{topbox}
    Analyze the provided Java code and identify the exact line numbers and statements that contain bugs. Return the exact buggy code lines and bug statements as they are without any explanations or code repair advice. Your response must strictly follow the format: 'Line X:bug statement'. Do not provide any additional information or suggestions.

    Example 1:

    Java code:

    [Java Source Code of Example 1]

    Response:

    [Defect Location in the Java Source Code of Example 1]

    Now, please check the following Java code to find out the potentially defective line numbers and bug statements:

    ...

    ...

    Java code:

    [Java Source Code for Fault Localization]
\end{topbox}

\begin{bottombox}
    bug report:

    [Bug Report for Java Source Code]
\end{bottombox}
\caption{Prompt template for Few-shot fault localization.}
\label{fig:prompt-Few-shot}
\end{figure}

\begin{figure}[htbp]
\centering
\begin{singlebox}
    You are a code defect localization expert specializing in Java. Your task is to read the provided Java code, identify potential errors, and locate the exact line number and buggy statement. Follow these steps:

    1. Read and understand the given Java code.

    2. Accurately number each line of the file to ensure that it is consistent with the line number of the original file.

    3. Examine variables, arrays, conditionals, loops, etc., in the code.

    4. Reason potential errors based on the code's possible issues. Focus on common error types such as array out-of-bounds, null pointer exceptions, division by zero, etc.

    5. According to the number in step 2, provide all possible error line numbers and error statements, and ensure that the line numbers are the same as those in the source file.

    Java code:

    [Java Source Code for Fault Localization]%

    Please begin your analysis and output the line number and buggy statement, without any further explanation, in the format: 'Line X: bug statement.'
\end{singlebox}
\caption{Prompt template for Chain-of-Thought fault localization on HumanEval-Java.}
\label{fig:prompt-cot-huameval-java}
\end{figure}

\begin{figure}[htbp]
\centering
\begin{singlebox}
    You are a code defect localization expert specializing in Java. Your task is to read the provided Java code, identify potential errors, and locate the exact line number and buggy statement. Follow these steps:

    1. Read and understand the provided bug report to extract symptoms, error types, and relevant clues (e.g., exception types, input conditions, or misbehavior).

    2. Read and understand the given Java code.

    3. Accurately number each line of the file to ensure that it is consistent with the original file line numbers.

    4. Examine variables, method calls, arrays, conditionals, loops, and other constructs that may relate to the symptoms described in the bug report.

    5. Based on the bug report and your understanding of the code, reason about potential causes — such as array out-of-bounds, null pointer exceptions, off-by-one errors, incorrect conditions, etc.

    6. Identify all possible buggy line numbers and their corresponding statements. Be precise — ensure that line numbers match the original source file.

    Java code:

    [Java Source Code for Fault Localization]%

    bug report:

    [Bug Report for Java Source Code]

    Please begin your analysis and output the line number and buggy statement, without any further explanation, in the format: 'Line X: bug statement.'
\end{singlebox}
\caption{Prompt template for Chain-of-Thought fault localization on Defects4J.}
\label{fig:prompt-cot-Defects4J}
\end{figure}